\documentclass[twocolumn]{aastex63}

\usepackage{amsmath}

\begin{document}

\title{Spectroscopic follow-up of Gaia exoplanet candidates:\\ Impostor binary stars invade the Gaia DR3 astrometric exoplanet candidates.}

\author[0000-0003-2173-0689]{Marcus L.\ Marcussen}
\affil{Stellar Astrophysics Centre, Department of Physics and Astronomy, Aarhus University, Ny Munkegade 120, 8000 Aarhus C, Denmark}
\email{E-mail: marcuslmarcussen@gmail.com}

\author[0000-0003-1762-8235]{Simon H.\ Albrecht}
\affil{Stellar Astrophysics Centre, Department of Physics and Astronomy, Aarhus University, Ny Munkegade 120, 8000 Aarhus C, Denmark}



\begin{abstract}

In this paper we report on the follow-up of five potential exoplanets detected with Gaia astrometry and provide an overview of what is currently known about the nature of the entire Gaia astrometric exoplanet candidate sample, 72 systems in total. We discuss the primary false-positive scenario for astrometric planet detections: binary systems with alike components that produce small photocenter motions, mimicking exoplanets. These false positives can be identified as double-lined SB2 binaries through analysis of high resolution spectra. Doing so we find that three systems, Gaia DR3 1916454200349735680, Gaia DR3 2052469973468984192, and Gaia DR3 5122670101678217728 are indeed near equal mass double star systems rather than exoplanetary systems. The spectra of the other two analyzed systems, HD 40503 and HIP 66074, are consistent with the exoplanet scenario in that no second set of lines can be found in the time series of publicly available high resolution spectra. However, their Gaia astrometric solutions imply radial-velocity semi-amplitudes $\sim$\,3 (HD 40503) and $\sim$\,15 (HIP 66074) larger than what was observed with ground based spectrographs. The Gaia astrometry orbital solutions and ground-based radial-velocity measurements exhibit inconsistencies in six out of a total of 12 exoplanet candidate systems where such data are available, primarily due to substantial differences between observed ground-based radial-velocity semi-amplitudes and those implied by the Gaia orbits. We investigated various hypotheses as to why this might be the case, and though we found no clear perpetrator, we note that a mismatch in orbital inclination offers the most straightforward explanation.

\end{abstract}

%\keywords{Exoplanets, binaries, Astrometry}


\section{Introduction} \label{sec:intro}

The detection of exoplanets through the astrometric reflex motion of their host stars has been predicted to revolutionize the field of exoplanetary science by increasing the total number of known exoplanets by a factor ${\sim}$\,4 after the nominal 5 yr mission of Gaia \citep{2016GaiaMission} and a factor ${\sim}$\,14 for the extended 10 yr mission \citep{perryman_astrometric_2014}. A benefit of the astrometric planet detection method over the Radial-Velocity (RV) method is that it provides the full orbital solution and the mass of the planet, though obtaining the precision needed for detecting planetary-mass companions is comparatively harder with this method. The astrometric detection sensitivity increases with orbital separation, opposite to the RV and transit methods which are most suitable for shorter period systems. This makes the methods highly complementary.

\begin{figure*}
    \centering
    \includegraphics[width = 1\textwidth]{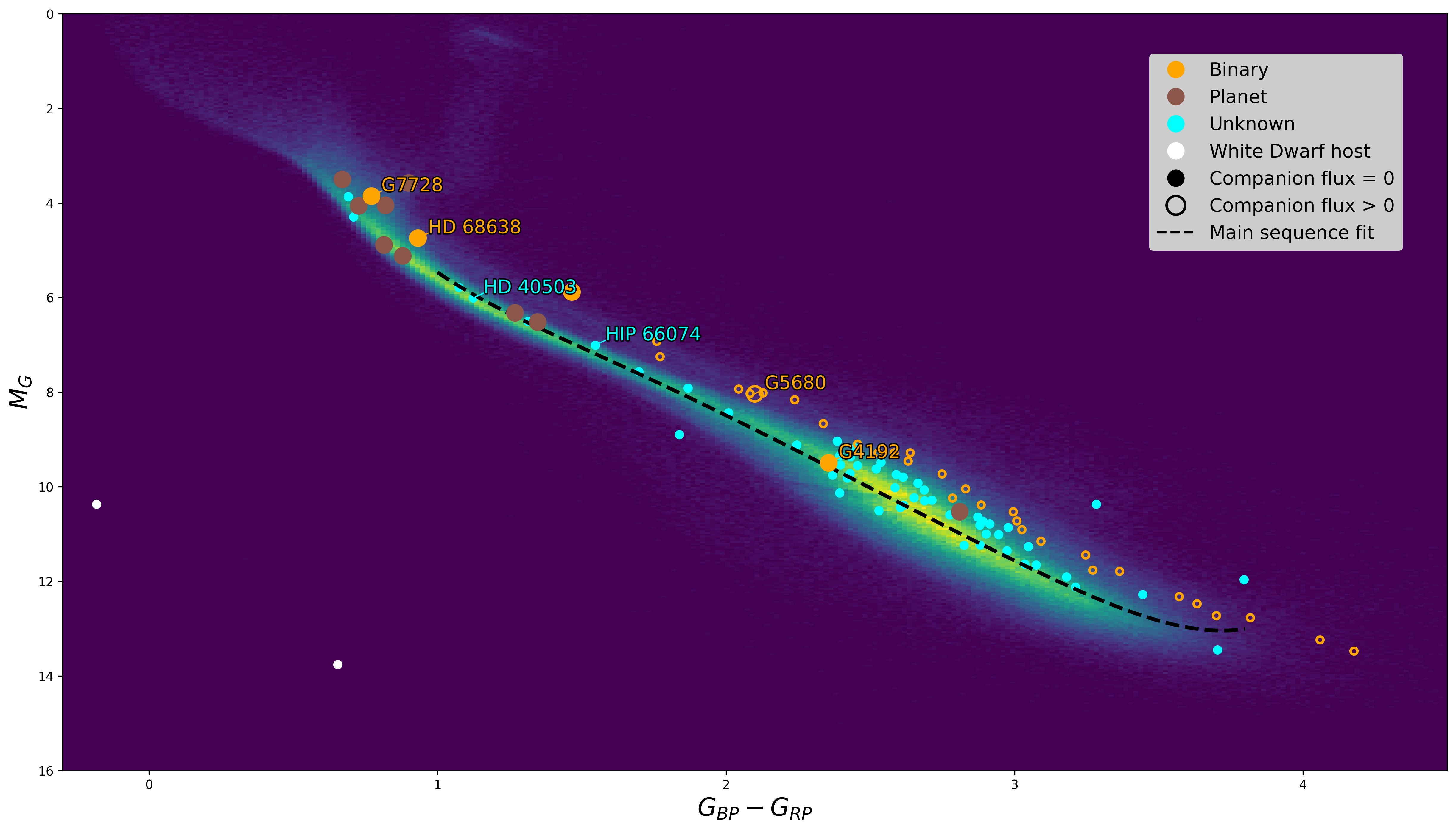}
    \caption{\textbf{Astrometric exoplanet candidate systems in an HR diagram.} All Gaia DR3 systems where the astrometric two-body motion of the source is small enough to indicate a companion with a mass $<20M_{\rm Jup}$, assuming the companion does not contribute any flux, are shown here as circles. The released Gaia astrometric exoplanet candidates are shown using closed circles while open circles indicate systems where a lower bound on the flux of the companion is non-zero as per the \texttt{binary\_masses} table. Large circles indicate systems where the nature of the companion is confirmed, either here or in the literature, see also table \ref{tab:results}. The systems are colored according to the current knowledge of the nature of the companion, where brown indicates a planetary companion, orange indicates a stellar companion and cyan indicates that it is unknown. The two white dots indicate two white dwarf hosts where the companion nature is unknown. The comparison stars shown in the background are the 1\,000\,000 nearest to us Gaia DR3 stars whose estimated parallax is more than 10 times its uncertainty. The density of these stars is represented as a purple-to-yellow heatmap, where purple (yellow) indicates a low (high) density. Since most of the planet candidates are too faint to have Gaia DR3 reddening and extinction parameters, candidates and comparison stars alike are shown using their uncorrected colors and magnitudes. The black dashed line represents the best fit to the Main Sequence. The six systems discussed in detail in section \ref{sec:examples} are indicated by their name.}
    \label{fig:HR_candidates}
\end{figure*}

On June 13 2022 the Gaia Data Release 3 (DR3) was published \citep{2022DR3}. This is the first data release where non-single source solutions are included. A single Keplarian astrometric orbit model is used. The 169\,277 orbital solutions released were compiled from 34 months of Gaia observation \citep{halbwachs_gaia_2022}, illustrating the potential of astrometric planet searches. However, the majority of these solutions stem from double star systems, which produce much larger astrometric signals than planetary systems. A small number of the released orbital solutions (1162) are from the more resource-intensive "exoplanet" pipeline \citep{holl_gaia_2022} aimed at low signal-to-noise ratio targets with substellar and particularly exoplanetary mass companions. The inputs for this exoplanet pipeline were either targets where the two body orbit model of the "binary" pipeline \citep{gaia_collaboration_gaia_2022-1} did not improve enough the single star fit (the \texttt{orbitalAlternative} solutions) or were pre-selected targets (\texttt{orbitalTargetedSearch}). Many of the pre-selected targets were chosen because they were known to harbor planets or brown dwarfs beforehand. In total 1843 astrometric brown dwarf and 72 exoplanet candidates were published in DR3 \citep{gaia_collaboration_gaia_2022-1}. 
%In order to keep the results self-contained \cite{gaia_collaboration_gaia_2022} did not use external information to validate the astrometric candidate systems, although a number of these systems have been studied before. 

\begin{figure*}
    \centering
    \includegraphics[width = 1\textwidth]{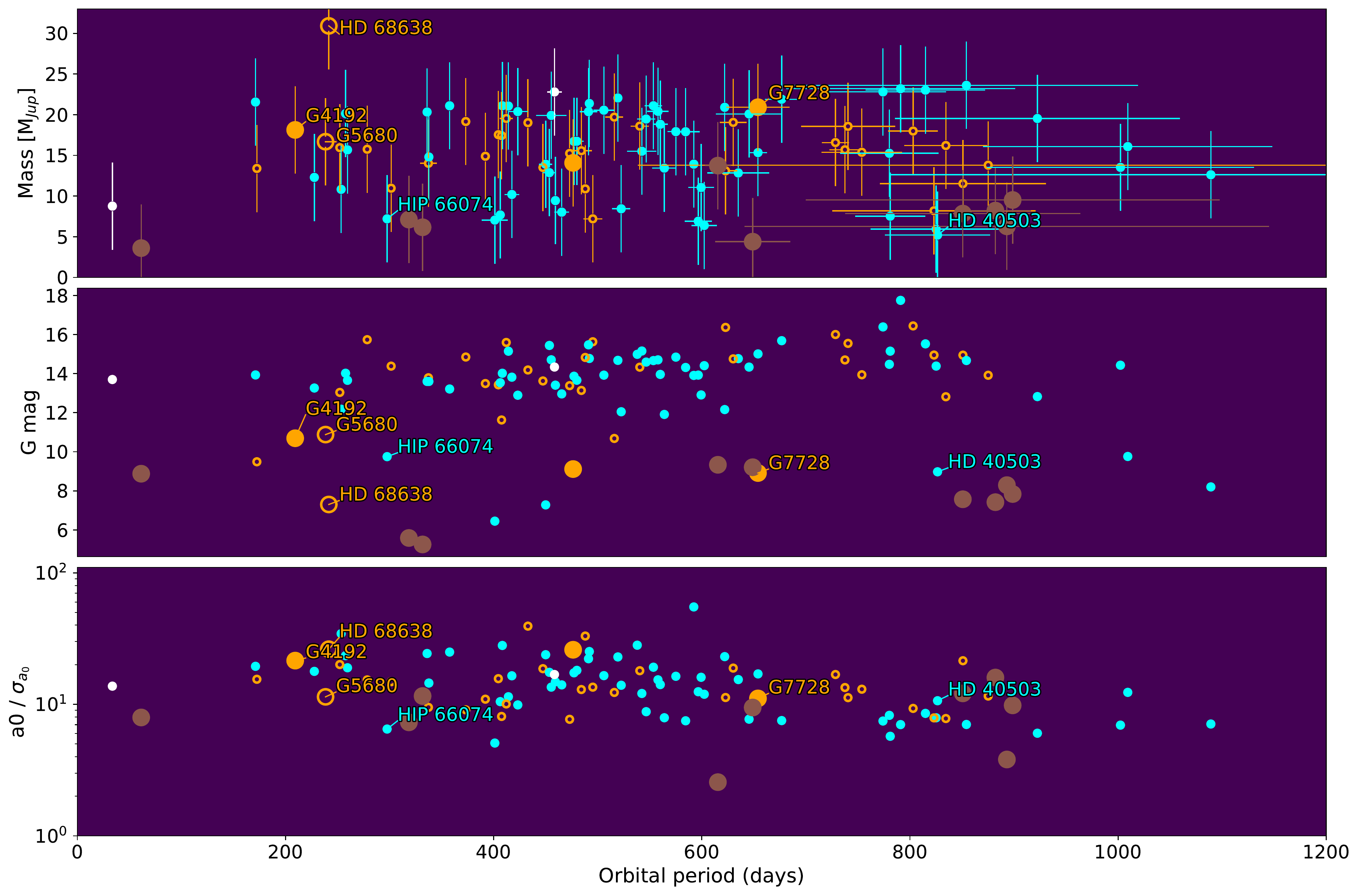}
    \caption{\textbf{Exoplanet candidate parameters.} 
    \textit{Top panel:}
     Companion masses of the same sample as shown in Fig.~\ref{fig:HR_candidates} are calculated under the assumption that their flux contribution is negligible compared to their host and assuming Gaussian uncertainties in photocenter semi-major axis a0, orbital period P and primary mass $M_1$. \textit{Middle panel:} The G magnitudes of the sample. \textit{Bottom panel:} The size of the photocenter semi-major axis relative to its uncertainty, which can be taken as a proxy for the robustness of the Keplarian orbit detection.
    }
    \label{fig:overview}  
\end{figure*}

An overview of all 102 systems with astrometric signals small enough to naively imply a companion mass $<20 M_{\rm Jup}$ are shown in Fig.~\ref{fig:HR_candidates} in a color-magnitude diagram. In order to get a sense of the prevalence of binary systems to exoplanetary systems, the figure includes 30 systems that are not among the 72 released exoplanet candidates \footnote{https://www.cosmos.esa.int/web/gaia/exoplanets}, due to their high likelihood of being binaries. For the masses of the primary stars we used the $m_1$ masses from the binary\_masses table where available. For a few systems where these were not available we used the single-source mass\_flame from the astrophysical\_parameters table. The systems are colored according to the current status of their nature, also available in table \ref{tab:results}. A subset of the systems in this table have already been studied for consistency between the astrometric and radial velocity data by \cite{holl_gaia_2022}, \cite{gaia_collaboration_gaia_2022-1} and \cite{winn_joint_2022}. We discuss their results in the sections of the specific systems. In Fig.~\ref{fig:overview} we show for these same systems the companion mass estimates, apparent magnitudes, and the relative uncertainty of their astrometric orbits. We note that only 17 of the systems have companion mass estimates below the often-used planet/brown dwarf limit of $13 M_{\rm Jup}$.

With only one previous discovery \citep{Salvador2022}, astrometry is not yet a well-developed technique for the detection of exoplanets. The method comes with its own set of unique challenges. In particular, the photocenter (the apparent position of the objects) motion of a binary system can be arbitrarily small, and thereby mimic a low-mass companion, as the flux ratio of the two stars approaches their mass ratio. In order for Gaia to achieve its full planet detection potential we need to understand better the nature of these challenges, their prevalence as well as provide paths to overcome them. In this work we discuss possible steps to confirm or reject the planet status of an astrometric candidate system using high resolution spectra. Specifically we are employing both publicly available spectra (of the three potentially new astrometric exoplanet discoveries discussed in \citeauthor{holl_gaia_2022} \citeyear{holl_gaia_2022} and \citeauthor{gaia_collaboration_gaia_2022-1} \citeyear{gaia_collaboration_gaia_2022-1}) as well as new FIES spectra (of another three exoplanet candidates) that we obtained from the Nordic Optical Telescope.\footnote{These spectra have been obtained through the \textit{Fast-Track} channel available at the observatory. http://www.not.iac.es/observing/proposals/}

Our paper is structured as follows: We first summarize important aspects of astrometric binary signals, section~\ref{sec:astrometric summary}, and discuss how to differentiate between genuine planet detections and false positives, section~\ref{sec:differentiating}. In section \ref{sec:examples} we analyze six exoplanet candidate systems, and discuss our findings in section~\ref{sec:conclusion}.

\section{Astrometric signal} \label{sec:astrometric summary}

A star and its companion orbit their common barycenter. If the companion is an exoplanet, i.e.~dark and low mass, the observed astrometric reflex motion of the source is small and coincides with the motion of the star. If the companion is another star though, i.e.~it has a higher mass and contributes flux, the actual motion of the primary star is then larger, but the observable, the photocenter of the system, is located between the two components. Its astrometric amplitude is smaller than the actual motion of the primary star and  can be arbitrary small given the system parameters. In this section we outline how the observed semi-major axis relates to the flux and mass ratios of the components in systems where the primary component has a single companion (subsection \ref{sec:binary systems}) and in systems where an inner pair of stars orbit an outer companion (subsection \ref{sec:triple systems}). 

The total astrometric motion of a star orbited by a single companion is the linear combination of the 5-parameter single source astrometric model and a Keplarian 7-parameter model. The single-source model consists of the two position parameters RA \& Dec, the proper motions, $\mu_{\rm RA}$ \& $\mu_{\rm Dec}$, and the parallax, $\varpi$. The Keplarian model consists of the orbital period $P$, time of perisastron passage $T_0$, eccentricity $e$, semi-major axis of the star's orbit $a_1$, the orbital inclination $i$, the argument of periastron $\omega$, and the right ascension of the ascending node $\Omega$. The latter four parameters are the Campbell elements that may also be expressed by the Thiele Innes coefficients A, B, F, G. For more information on the Thiele Innes coefficients and how they are derived from the Gaia along-scan measurements, see \cite{gaia_collaboration_gaia_2022-1}. 

\subsection{Binary systems}
\label{sec:binary systems}
\begin{figure*}
    \centering
    \includegraphics[width = 0.95\textwidth, trim = {2cm, 0.2cm 2cm 2cm}]{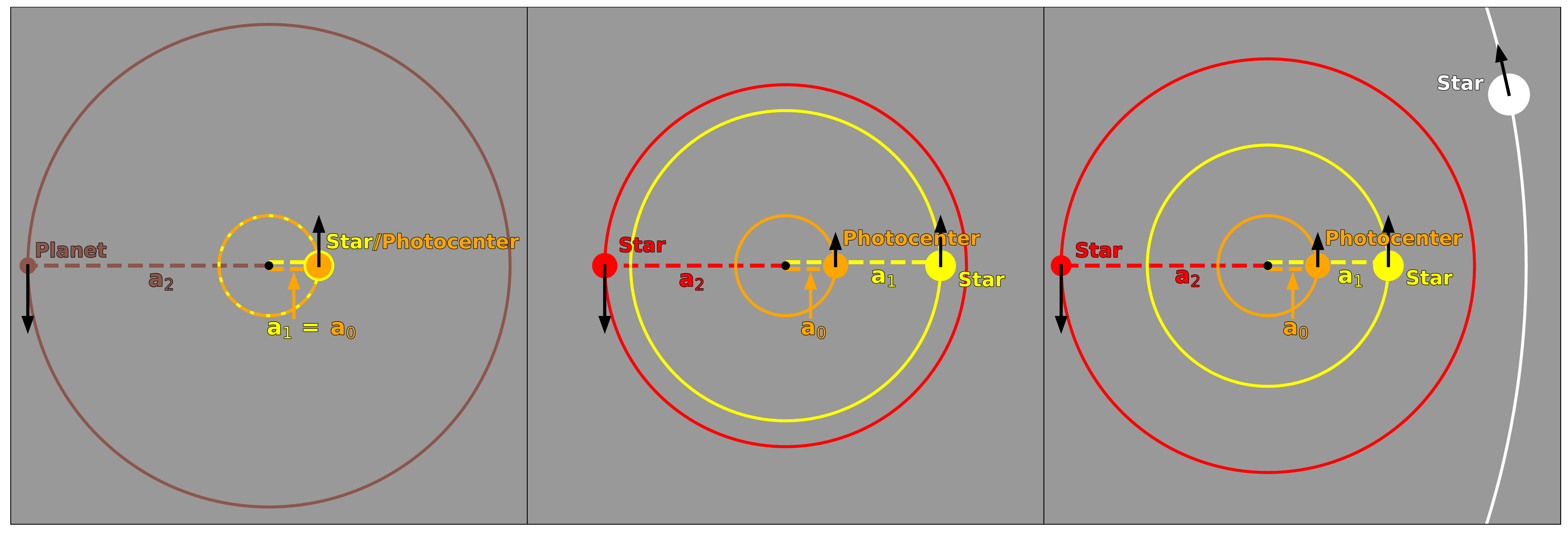}
    \caption{\textbf{Photocenter.} Not-to-scale illustration of three different geometries of face-on, circular orbits that the same astrometric signal may originate from. {\it Left panel:} An exoplanet orbiting a star. The semi-major axis of the exoplanet orbit, $a_2$, is much larger than the orbit of the star, $a_1$, due to the large difference in mass. The flux contribution from the planet is negligible and thus the photocenter closely follows the star.
    {\it Middle panel:} Twin stars (i.e. almost equal flux and mass stars) orbit their
    barycenter (black dot). The primary component, the yellow star, is slightly more massive than the secondary component, the red star, leading to a smaller semi-major axis $a_1$ than $a_2$. The ratio of the masses is slightly higher than the ratio of the received fluxes, $M_2 / M_1 > F_2 / F_1 $, resulting in a small photocenter semi-major axis $a_0$ that traces out (a diminished version of) the orbit of the primary component. {\it Right panel:} Two stars with significantly different masses and fluxes in a hierarchical triple system, where the center-of-mass of the red and yellow star orbit around a larger and brighter third component that is very far away and shown in white. The flux of the third component diminishes the size of the photocenter semi-major axis but does not contribute to the astrometric signal due to its very long period. The photocenter semi-major axis would be larger and more closely follow the yellow star if not for the third component.}
    \label{fig:photocenter}
\end{figure*}

For systems with a single companion, the semi-major axis of the relative orbit $a$ is the sum of the semi-major axes $a_1$ and $a_2$ around the barycenter of the star and companion respectively, see Fig.~\ref{fig:photocenter}. The ratio of the semi-major axes is equal to the inverse ratio of the masses, $M_2/M_1 = a_1/a_2 \equiv q$, giving

\begin{equation}
     a_1 = a \frac{M_2}{M_2+M_1} = a \frac{q}{q +1}\,.
\end{equation}

However, in completely unresolved systems where the secondary component has a non-negligible flux, we observe the motion of system's photocenter, i.e.~the flux-weighted mean position of the two components. This gives the following definition for the semi-major axis of the photocenter $a_0$:

\begin{equation}
\label{eq:PhotocenterAmplitude}
     a_0 = \left| \frac{F_1 a_1 - F_2 a_2}{F_1+F_2} \right| .
\end{equation}

Here $F_i$ indicate the flux of the i'th component (in the band-pass in which the astrometric signal is measured). Trivially, the astrometric observable, $a_0$, is equal to $a_1$ when the companion is dark ($F_2 = 0$). This scenario is illustrated in the left panel of Fig.~\ref{fig:photocenter}. A ratio of the fluxes of $\epsilon \equiv F_2 / F_1$ equal to one leads to $a_0=a_1-a_2$. If $q$ is also $1$, then the photocenter remains in the barycenter of the system. In the case that $F_2$ is non-zero the semi-major axis of the photocenter is given by:

\begin{align}
\label{eq:a0}
     a_0  &= \left|  \frac{F_1 a_1 - F_2 a_2}{F_1+F_2}\right| = \left| \frac{a_1 - \epsilon a_2}{1+\epsilon} \nonumber\right| \\
     &=\left|  \frac{a - a_2 - \epsilon a_2}{1+\epsilon} \nonumber\right| \\
     &= \left| \frac{a}{1 + \epsilon} \left( 1 - \frac{1}{1+ q} \left( 1 + \epsilon \right)\right) \nonumber\right| \\
     &= \left| a \left( \frac{1}{1+ \epsilon } - \frac{1}{1+ q}   \right) \right| . 
\end{align}

And making use of Kepler's third law we find:

\begin{align}
     a_0^3   &=\left| GM_1 \left(\frac{P}{2 \pi}\right)^{2} \left( 1+q \right)  \left( \frac{1}{1+ q } - \frac{1}{1+ \epsilon}   \right)^3 \right| \,.  
     \label{eq:a0_kepler}
\end{align}

The gravitational constant is indicated by $G$. 
A tiny photocenter movement may thus originate from two common but astrometrically indistinguishable scenarios: I) A single star orbited by a dark, low-mass companion or II) A double-star system with $\epsilon \approx q$, illustrated in the left and middle panels of Fig.~\ref{fig:photocenter} respectively.
%where $\epsilon$ is close to $q$.
Therefore in order to confirm an astrometric exoplanet candidate detected by the small astrometric reflex motion of its host star, additional measurements using different methods are needed. %to lift the degeneracy between these scenarios.

We note that if $q$ exceeds $\epsilon$, the location of the photocenter is between the barycenter and the \textit{secondary} component rather than the primary. This will result in a $180^\circ$ phase-shift in $\omega$ of the photocenter orbit. With RV measurements of the primary component, one would in principle be able to distinguish between $q > \epsilon$ and $q < \epsilon$ by the opposing sign of the RV signal to that inferred by the astrometric photocenter motion. However, because the astrometric motion is the sky-projection of the three-dimensional orbit one cannot a priori distinguish between the configurations ($\Omega$, $\omega$) and ($\Omega + 180^\circ$, $\omega + 180^\circ$). Even though we can resolve this ambiguity of the photocenter orbital configuration by combining RV data with an astrometric solution, we thus still face uncertainty about whether the orbit corresponds to the orbit of the primary or secondary component of the system. In other words, if the RV signal's sign contradicts our expectations from the photocenter orbit, it remains unclear whether this discrepancy arises from an incorrect assumption about $\omega$ or from a larger value of $q$ compared to $\epsilon$. In order to truly lift this degeneracy, second order effects like the light-travel time (light has a shorter travel distance when the component is on "our side" of the sky-plane) and local perspective effects (the parallax is larger when the component is on "our side" of the sky-plane) would need to be taken into account \citep{halbwachs_local_2009}. These effects are not modeled in Gaia DR3. If we do not know whether the photocenter orbit traces out the orbit of the secondary or primary, i.e.~if $q$ is larger or smaller than $\epsilon$, and if we have no prior on the flux ratio of the components, i.e.~$\epsilon$ is anywhere between 0 and 1, knowledge of $a_0$ alone is not enough to determine $M_2$.

\subsection{Triple systems}
\label{sec:triple systems}
Another configuration that could conceivably produce a small photocenter semi-major axis, $a_0$, is a hierarchical triple star system where a large amplitude astrometric signal from a double star system is muted by the presence of a brighter third component further away. If the distance to the third component is sufficiently large (but not so large as to be resolvable), its long term contribution to the astrometry would be undetectable in the Gaia data but the amplitude of $a_0$ would be diminished by a factor

\begin{equation}
    (F_1+F_2)/F_3.
\end{equation}

Where $F_3$ denotes the flux of the third star in the relevant band pass. This is illustrated in the right hand panel of Fig.~\ref{fig:photocenter}. The third component might also be a non associated source i.e., a chance alignment. 

\section{Differentiating between stellar and planetary companions} \label{sec:differentiating}

In this section we describe ways to determine the nature of a companion given the detection of an astrometric Keplarian signal with a small photocenter. 

\subsection{Color-magnitude relation}

The luminosity of a star on the Main Sequence can be well constrained by its effective temperature. A system of two identical, unresolved stars has twice the luminosity of a similar single star but the same effective temperature, in turn elevating its position relative to the single star Main-Sequence in an HR-diagram. This fact allows for an estimate of the flux ratio $F_1/F_2$ using isochrones \citep{gaia_collaboration_gaia_2022-1} giving a first indication of whether an astrometric two-body orbit is caused by a stellar or substellar companion. In Fig.~\ref{fig:HR_candidates} we can observe how all the systems whose flux ratio have been estimated to be non-zero this way (indicated by open circles), are located clearly above the single star Main Sequence, i.e. their luminosity is higher than expected, given its observed color, if the system were a single star. For DR3 systems with both astrometric and spectroscopic solutions, first $M_2$ is derived from the RV semi-amplitude, K, via equation 3 in \cite{gaia_collaboration_gaia_2022-1}, and then the flux ratio is inferred from how much smaller the astrometric signal is (our equation \ref{eq:a0}). The masses shown in Fig.~\ref{fig:overview} are all calculated using eq.~\ref{eq:a0_kepler} from the astrometric semi-major axis alone, i.e.~using $F_2 = 0$ for all systems. We do this in order to get a better sense of the ratio of stellar companions to substellar companions among all orbital solutions with small $a_0$.

\subsection{Spectra}
High resolution spectroscopic follow-up observations can be used to reveal the impostor binaries among the astrometric exoplanet candidate systems. Spectra from star-planet systems are fundamentally different from star-star systems: only one set of absorption lines are present and the radial velocity of the host star is much smaller due to the lower mass of the companion. This means the Doppler-shift of the spectrum over time is much smaller. Depending on whether an impostor binary system consists of two truly identical twins (i.e.\ have very similar effective temperatures, luminosities, rotation speeds and masses) or are significantly different stars whose $q$ happen to be close to $\epsilon$ (e.g.\ a red giant and a Main Sequence star), the two sets of oppositely shifting absorption lines produced by the binary system may present themselves differently:

\paragraph{Uneven twin impostors } 
~If one of the component is of a very different stellar class and/or has a very different rotation rate, the absorption lines from a single component may dominate the spectrum of the system. Such a system would be classified as a single lined binary (SB1) system. Impostors of this type allow us to measure the orbital RV semi-amplitude of the component most visible in the spectra, which will be significantly larger than if it were an exoplanetary system. If an astrometric exoplanet candidate is in fact an SB1 binary, a few observations at different orbital phases are enough to identify it as such. If the orbital period and time of periastron passage are well determined from the astrometry such that the follow-up observations are ensured to be undertaken at advantageous points of the orbit, just two observations may be enough.

\paragraph{Identical twin impostors }
~If, on the other hand, the two stars are very alike in every way, their absorption lines will be similar in shape and size. Additionally, the RV shift of each stars absorption lines will be equal in amplitude but with opposite signs. These impostors may be identified as double lined (SB2) binaries rather than exoplanets with a single observation, if the RV separation of the two stars is of the order of (or larger than) the combined absorption line-width of the stars and the spectral resolution is sufficient. In that case, their identification  as false positive candidates is trivial.

However, depending on the spectral resolution of the instrument, the orbital phase of the observations, the orbital inclination, the period, the masses of the components and the absorption line-widths, each stars absorption lines may not always be clearly separated. When that is the case, the RV shift of the two sets of absorption lines may be identified as a change of the line width, i.e., the FWHM (Full Width at Half Maximum), of the {\it combined} line. Observing a small RV semi-amplitude of what appears to be a line of a single star does therefore not rule out the possibility that the companion is an identical stellar twin. The midpoint of this combined line from two unresolved sets of absorption lines will RV shift slightly, given by small flux and/or rotation speed differences of the two stars. We simulated the RV shift of a combined line from two almost identical twin binaries to test whether its small RV semi-amplitude could mimic that of a star orbited by a planetary companion.

%We varied the $\epsilon/ q$ ratios of the simulated systems from $0.95$ to $1.05$. Our results showed that for all configuration of $M_1$, $M_2$, $F_1$ and $F_2$ the RV semi-amplitude of the combined line is indeed diminished to a very similar degree (within a few percent) to that of the astrometric signal, as long as the rotation speeds of the two components did not also differ significantly. In other words, if the spectral resolution is low, both the astrometric and RV data may falsely -- but consistently between the data sets -- indicate the presence of an orbiting exoplanet. In those cases, changes in the line width will be a better indicator for identifying the stellar nature of the companion.

\begin{figure*}
    \centering
    \includegraphics[width = \textwidth]{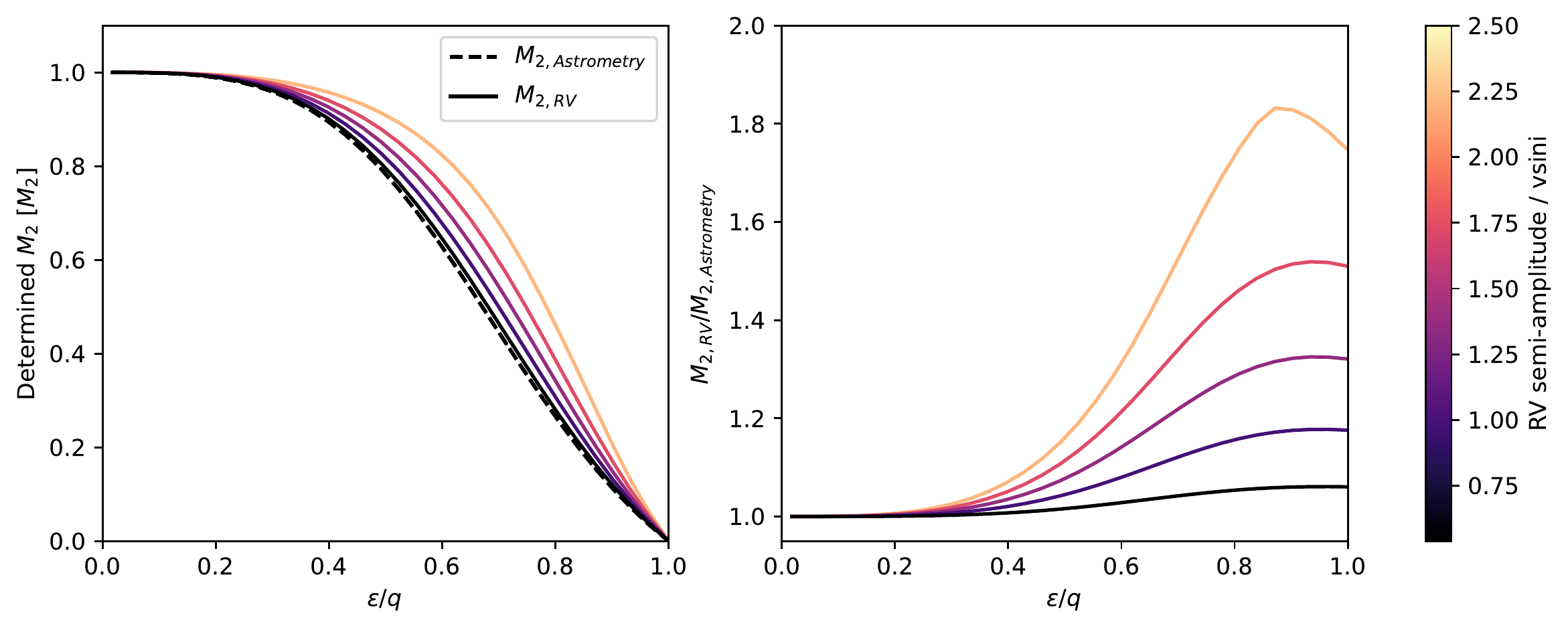}
    \caption{\textbf{Astrometry and RV simulations. }\textit{Left:} The mass of a companion as determined by astrometry (dashed line) and RVs (solid lines) relative to its actual mass as a function of the flux ratio divided by the mass ratio, $\epsilon / q = (F_2 / F_1) / (M_2 / M_1)$. The line colors indicate the ratio of the RV semi-amplitude to the absorption line widths, $v \sin i$. \textit{Right:} The ratio of the RV determined mass to the astrometry determined mass.}
    \label{fig:m2s}
\end{figure*}

Fig.~\ref{fig:m2s} illustrates this. Here we assumed a simple $\epsilon = q^4$ mass-luminosity relation, though as long as the mass ratio is always closer to $1$ than the flux ratio, the steepness of the relation did not qualitatively change our results. We then varied the $\epsilon / q$ ratios of the simulated systems from $0$ to $1$ and calculated the mass of the companion as inferred by the astrometric method and the RV method, see the left panel of Fig.~\ref{fig:m2s}. The width of the simulated absorption lines of the primary and secondary components were kept equal. In the right panel we show that as long as the RV semi-amplitude is $\lesssim$ $v \sin i$ of the components, the determined companion mass as determined by the two methods is very similar. However, the ratio of the RV determined mass to the astrometric mass, $M_{2, RV} / M_{2, astrometry}$ becomes increasingly sensitive to small changes in rotation speed differences between the components and in the mass ratio of the components as $\epsilon / q$ approaches $1$. In conclusion, if the spectral resolution is low, both the astrometric and RV data may falsely -- but consistently between the data sets -- indicate the presence of an orbiting exoplanet for some systems. In those cases, changes in the line width will be a better indicator for identifying the stellar nature of the companion.

The Gaia spectral line broadening parameter \texttt{vbroad}, which is the median of epoch vbroad measurements from the Gaia Radial Velocity Spectrometer, may be useful for identifying twin binary systems without the need for ground-based follow-up. 
%This is evident in the statistics of the Gaia DR3 data where 
The median \texttt{vbroad} of the 412 Gaia non-single-source solutions with flux ratios above 0.25 is 15.9~km\,s$^{-1}$, whereas it is 10.9~km\,s$^{-1}$ for the 19\,799 systems with flux ratios below 0.25. Unfortunately, \texttt{vbroad} is only available for stars brighter than G$_{\rm mag} 12$ and most current astrometric exoplanet candidates are found around fainter stars (see Fig.~\ref{fig:overview}). 

\begin{figure*}
    \centering
    \includegraphics[width = \textwidth, trim = {4cm, 0.2cm 3cm 2.8cm}]{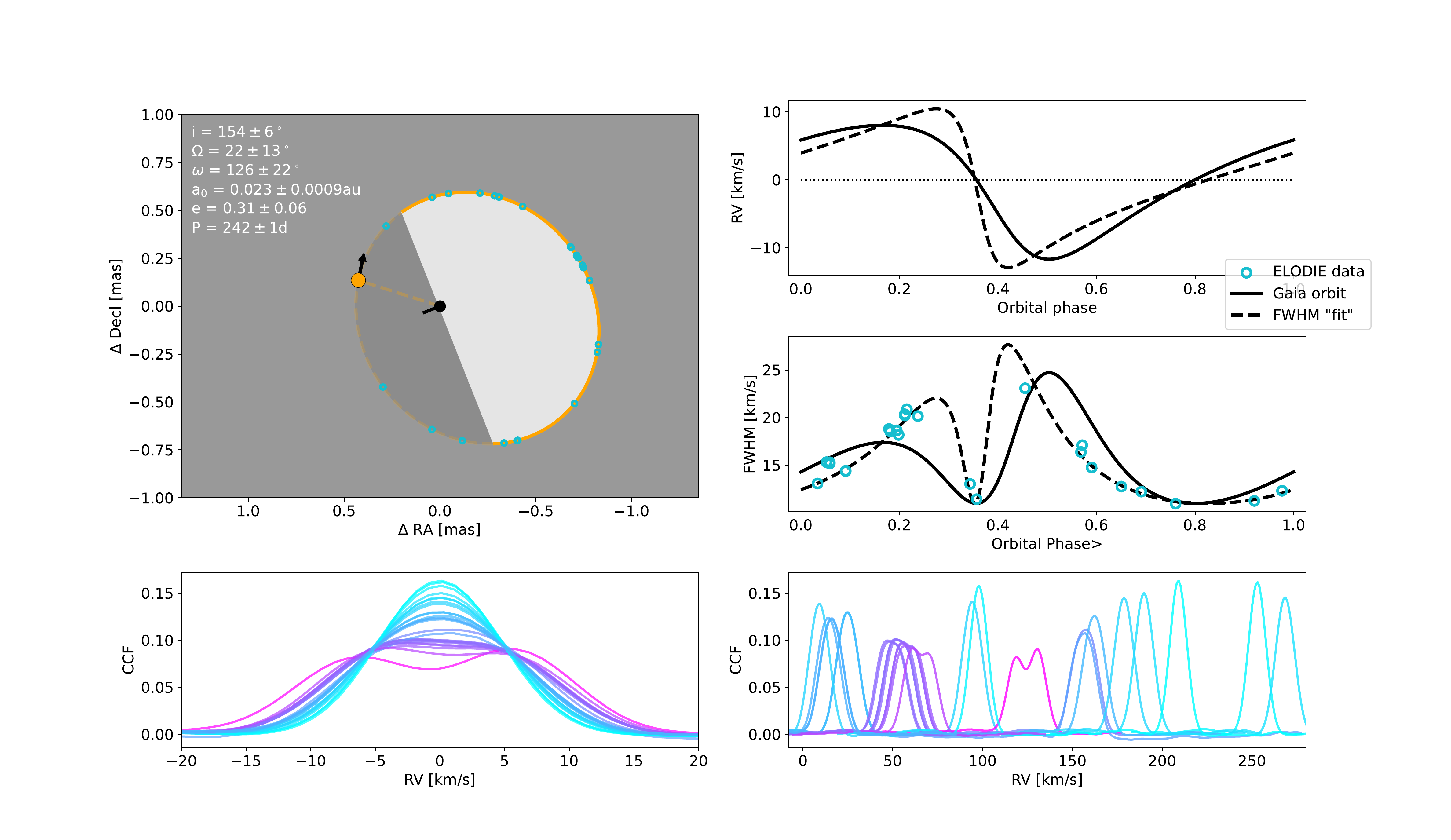}
    \caption{\textbf{HD 68638.} \textit{Top left:} The geometry of the Gaia two-body orbital solution for the photocenter is shown with north pointing up and east pointing left. The Gaia parameters are printed as well. The barycenter is shown as a black dot and the star is shown at periastron as an orange dot with the direction of motion indicated by the black arrow. The line through the barycenter represents the orbital angular momentum vector. The positive direction of the angular momentum is marked with an arrow that is occluded by the orbital disk in this particular view. The light gray part of the orbit with a solid orange outline indicates the near side of the sky plane and the dark gray part with a dashed outline indicates the far side of the sky plane. The blue circles are the location on the orbit of the RV measurements. \textit{Bottom left:} CCFs from all 27 publicly available ELODIE spectra are shown %\footnote{Available at https://dace.unige.ch/radialVelocities/?pattern=HD68638A} 
    and colored according to their FWHM. \textit{Top right:} The solid black line indicates the expected radial velocity curve of the primary component based on the Gaia-derived orbital and stellar parameters assuming two equal-mass companions. The dashed line represents an orbit with a higher eccentricity of $0.6$. \textit{Middle right:} The FWHM of the CCFs are shown as blue dots as a function of orbital phase. The solid black line is the FWHM as a function of orbital phase inferred from the Gaia orbit. The dashed black line represents the FWHM of a more eccentric orbit that is a better fit to the observed FWHM. \textit{Bottom right:} CCFs plotted with a horizontal offset so their midpoints are located under their respective FWHM data points of the middle right panel.}
    \label{fig:HD68638}
\end{figure*}

\section{Example systems} \label{sec:examples}

In this section we analyze spectra of six astrometric exoplanet candidate systems to clarify the nature of the companions that cause a Keplarian motion of their host star.
%In this work we investigate six systems in detail whose astrometric signal is compatible with potential new exoplanet discoveries.
The first three systems, HD 68638, HD 40503 and HIP 66074, also highlighted by \cite{holl_gaia_2022}, have publicly available archival high resolution spectra that we reanalyzed. The latter three systems, Gaia DR3 2052469973468984192, Gaia DR3 1916454200349735680 and Gaia DR3 5122670101678217728, were investigated using high resolution spectra that we obtained with the FIES spectrograph installed at the Nordic Optical Telescope (NOT). The RVs inferred from these spectra enable us to supply the non-degenerate orbital solutions for the photocenters of the former three systems. Below we discuss our findings for each system separately.

\subsection{HD 68638}

HD 68638 (Gaia DR3 1035000055055287680) is a G magnitude $7.3$ system whose orbital solution was highlighted in \cite{holl_gaia_2022} and \cite{gaia_collaboration_gaia_2022-1} because of its small photocenter semi-major axis. Despite its slightly elevated position in the HR diagram (Fig.~\ref{fig:HR_candidates}), its Gaia DR3 lower bound flux ratio is $0$. Although this system has previously been characterized as a binary system \citep{Busa2007}, in order to test our methodology and to better understand the characteristics of false-positives, we reanalyzed the 27 publicly available ELODIE spectra of this system. In the bottom left panel of Fig.~\ref{fig:HD68638} we present the CCFs (Cross-Correlation Functions) obtained from these spectra along with the Gaia astrometric two-body orbital solution in the top left panel. Where the relative RVs are largest the CCF displays two peaks with similar widths and heights, confirming the twin SB2 status of this system. The same conclusion was reached by \cite{holl_gaia_2022} and \cite{gaia_collaboration_gaia_2022-1}. In the top right panel of the same figure we show the RV curve implied by the Gaia astrometric solution. To illustrate how the FWHM correlates with orbital phase, we show in the middle right panel the expected FWHM of the average absorption line as implied by the Gaia orbit. The measured FWHM of the ELODIE spectra are superimposed on this. By increasing the eccentricity of the Gaia orbit (shown as a dashed line) from 0.32 to 0.6 we get a better fitting orbit solution with the observed FWHM. For the sake of consistency we calculated the FWHM of the total CCF even when the CCF splitting is apparent. 
%Given that this is one of the brightest exoplanet candidate systems, one would have expected this system to have a Gaia SB1 or SB2 solution. Most likely though, with the limited resolution ($\lambda / \delta \lambda$) of the Gaia Radial Velocity Spectrometer the spectrum only shows a single set of almost static absorption lines, where the first fact excludes an SB2 classification and the second fact excludes SB1.

\subsection{HD 40503}

\begin{figure*}
    \centering
    \includegraphics[width = \textwidth, trim = {4cm, 2.2cm 3cm 2.8cm}]{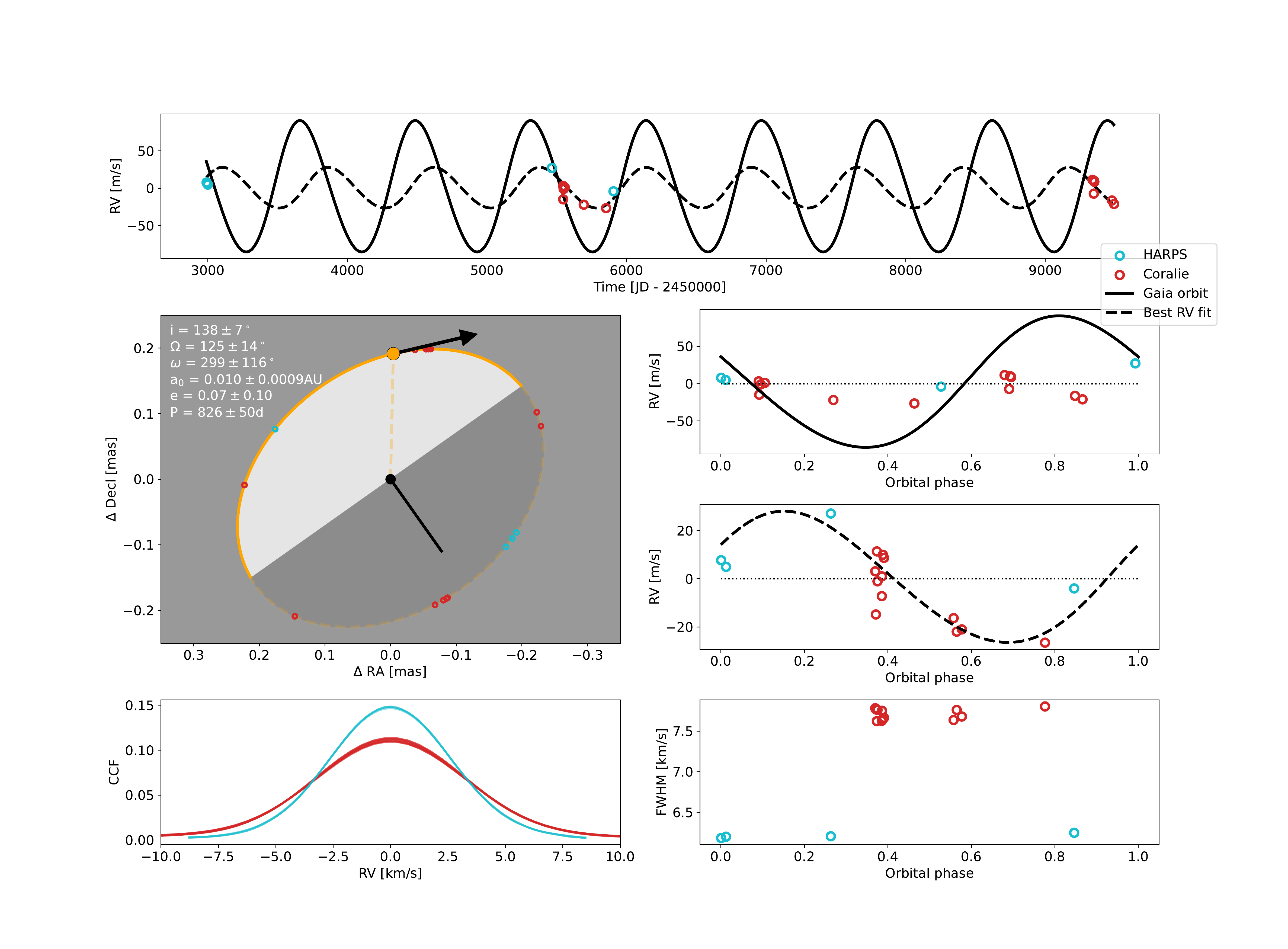}
    \caption{\textbf{HD 40503.} Shown here is the geometry and orbital parameters of HD 40503 as well as RV measurements in red and blue from the HARPS and CORALIE spectrographs respectively. The black dashed line represents the best fit to the RV data alone.}
    \label{fig:HD40503}
\end{figure*}

HD 40503 (Gaia DR3 2884087104955208064, HIP 28193) was discussed in \cite{holl_gaia_2022} and \cite{gaia_collaboration_gaia_2022-1} as a potential new exoplanet discovery, because the $826$d orbital period derived by Gaia is consistent with the periodicity in publicly available Coralie and HARPS spectra. In Fig.~\ref{fig:HD40503} we show CCFs %reobtained by us 
from these spectra as well as the FWHM as a function of orbital phase, none of which indicate binarity. Because we see no indications that the companion is a binary, we compare the Gaia RV curve to the HARPS and CORALIE data points (top and right panels of Fig.~\ref{fig:HD40503}). Although we also find the Gaia period of $826 \pm 50$\,d to be consistent within $2\sigma$ of our $758$\,d best fit to the RV data, the RV semi-amplitudes are highly inconsistent: The Gaia photocenter semi-major axis of $a_0 = 0.010$~au ($0.25$~mas) implies a companion mass of $5.18 \pm 0.59\,M_{\rm Jup}$, assuming $F_2 = 0$, whereas the observed RV amplitude would imply a companion mass of $1.55\pm 0.18\,M_{\rm Jup}$. The observed discrepancies are possibly due to stellar activity affecting the RVs as noted by \cite{holl_gaia_2022} and \cite{winn_joint_2022}. However, it is interesting that the grounds-based RVs of HIP 66074 (see also below) as well as HR 810 \citep{winn_joint_2022} also have much smaller RV semi-amplitudes than their Gaia orbit counterparts.%While it is not possible to rule out a planetary-mass companion of this system, we would argue that due to the RV amplitude discrepancy, the CORALIE RV data is insufficient to validate the Gaia orbit after all. 

\subsection{HIP 66074}
\label{sec:HIP66074}
\begin{figure*}
    \centering
    \includegraphics[width = \textwidth]{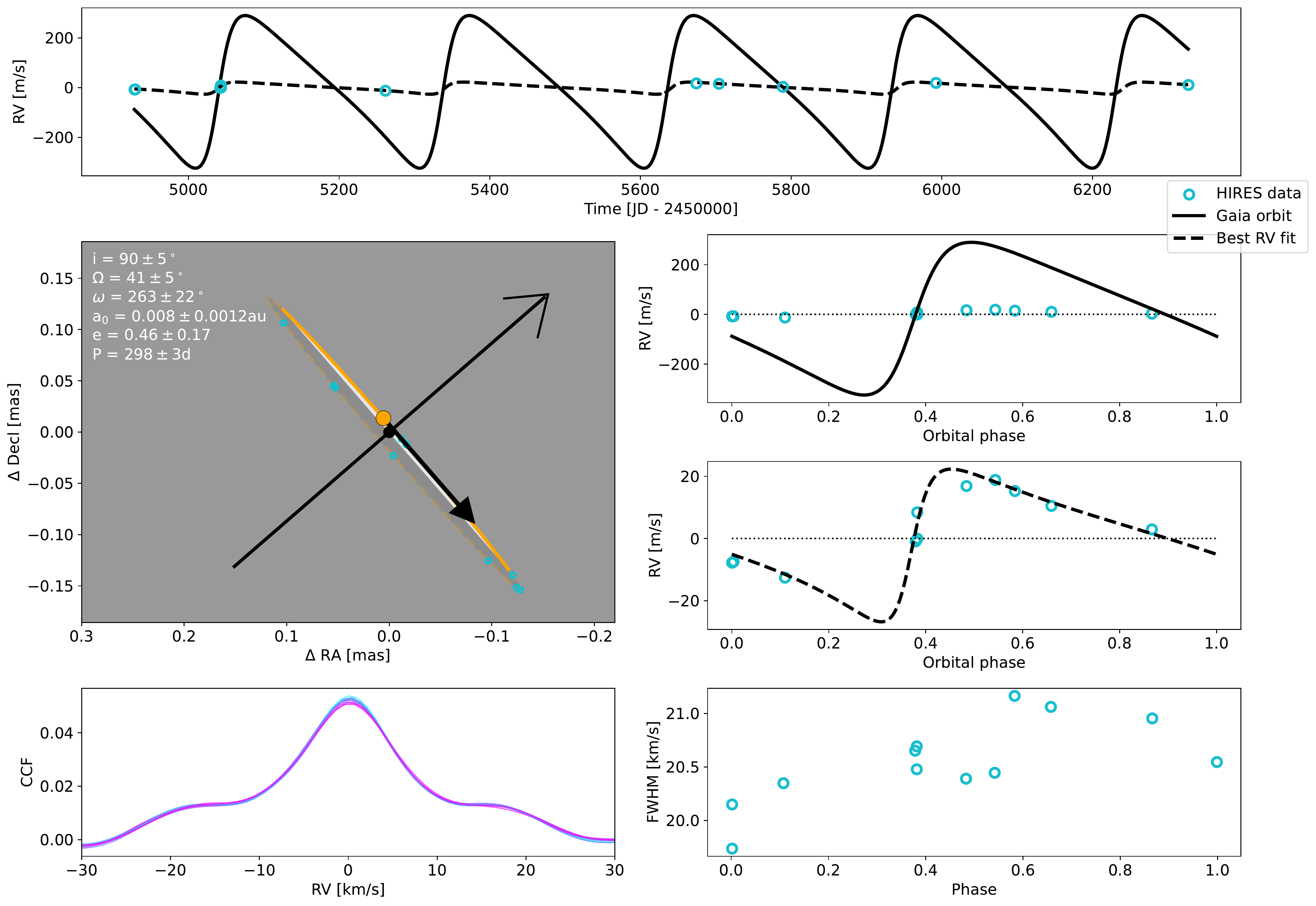}
    \caption{\textbf{HIP 66074.} Shown here is the Gaia geometry and orbital parameters of HIP 66074 as well as RV measurements in blue derived from HIRES spectra. The Gaia astrometry and RV data are in excellent agreement except for a large factor 15 disagreement in the RV amplitudes between the data sets. In the bottom left we show CCFs that we obtained from the HIRES spectra. In contrast to Fig.~\ref{fig:HD68638}, the CCFs from the different observations are so similar that they look like one, ruling a stellar companion.}
    \label{fig:HIP66074}
\end{figure*}

The system HIP 66074 (Gaia DR3 1712614124767394816) was discussed by \cite{holl_gaia_2022} and \cite{winn_joint_2022} because of its small photocenter semi-major axis of $0.0075$~au (0.21 mas), which is consistent with a $6.9\pm 1.1 M_{\rm Jup}$ mass companion. In Fig.~\ref{fig:HIP66074} we show our CCFs derived from the public HIRES spectra that show no sign of double lines or FWHM changing with orbital phase, ruling out a twin binary system.

The orbital period, eccentricity, time of periastron passage, and argument of periastron from the Gaia orbital solution and the best RV fit to the ground-based data are all within just $1\sigma$ of each other \citep{winn_joint_2022}. However, there is an enormous factor 15 semi-amplitude discrepancy between the ground-based RV data and Gaia astrometric data as highlighted by \cite{holl_gaia_2022} and \cite{winn_joint_2022}. Interestingly, although they caution its interpretation, \cite{winn_joint_2022} do find an excellent fit between the two data sets with the only caveat that the companion needs a substantial flux ratio of $\epsilon = 1 \pm 0.12 \%$. This flux ratio is well above the mass ratio of $q = 0.063 \%$ which would imply that the photocenter orbit is a scaled-down version of the companion's orbit and not the host star's orbit. It also implies a seemingly unrealistic luminosity of the planet: assuming the companion has a radius of $1 R_{\rm Jup}$, radiating 1\% of the flux seen from HIP 66074 as a blackbody puts its effective temperature at around $3\,400$~K! To find more plausible explanations to the large semi-amplitude discrepancy, we considered several hypotheses. Below we go through these and discuss the challenges each has:

\paragraph{There is no companion }~The independent detections of an orbit with similar orbital parameters by Gaia astrometry and HIRES RVs makes it very unlikely that the Keplarian motion of the source is a simple case of a false-positive detection. 

\paragraph{Stellar activity affecting the RVs }
~Perhaps, as may be the case for HD 40503, stellar activity affects and mutes the RVs? If that were the case, we would not expect the RV semi-amplitude to be the \textit{only} affected parameter. We might also expect deformations in the CCFs, which we do not see (Fig.~\ref{fig:HIP66074}).

\paragraph{Additional planets }
~Would affect the reflex motion of the host star and thus the RVs and could erase or mute a two-body orbital signal. However, because the astrometric and RV data would be affected equally by additional planets, it does not describe the large observed RV semi-amplitude difference. Additionally, if the motion of the host star is the linear combination of several Keplarian orbits, the simple two-body model would not fit the data as well as it does.

\paragraph{The Gaia semi-major axis is overestimated }
~If the actual size of $a_0$ was a factor 15 smaller than the reported value of $0.21 \pm 0.03$\,mas then the two independent data sets would agree. The issue with this explanation is that an astrometric two-body orbit with an angular diameter of $0.014$\,mas would then have been detected, which is well below the precision of Gaia.

\paragraph{Triple system }
~Another potential explanation is that we are observing the orbit of a binary system which itself is part of a hierarchical triple system (Fig.~\ref{fig:photocenter}, right panel). In such a system the astrometric and RV signature of the binary orbit could be muted by different amounts, preserving all observed characteristics of the orbit other than their apparent amplitudes.
The mechanism is as follows: The hypothetical spectrum from such a system would be a combination of lines from all three components, but only the spectra from the two binaries would shift in velocity space. If the components are not separately identified in the spectrum, the obtained {\it RV signal} would be muted relative to a signal from just the binary system. The \textit{astrometric} signal of the close binary would also be muted by the polluting light from the third component, but not necessarily by the same amount as the RV signal. As shown in section \ref{sec:astrometric summary}, the astrometric signal is muted by a factor $(F_1 + F_2)/F_3$ by a bright, third component on a long orbit.

The boundary conditions of this scenario are i) the third component must be on a distant enough orbit that the corrections to the two-body orbit of the double-stars is undetectable and ii) it is close enough to the binary system that it cannot be spatially resolved by Gaia. To get an estimate of the possible range of orbital periods where both of these criteria are fulfilled, we conservatively assume a total mass of the triple system equal to the \texttt{binary\_masses} $m_1$ mass. This puts an upper limit on the orbital period of the third component's orbit of around 16.5 years. To calculate this, we used its distance to us of 35 pc and Gaia's 0.23 arcsec two-body resolving limit\footnote{https://www.cosmos.esa.int/web/gaia/science-performance}. On the near-end the orbital period would need to be several times the 34 month baseline of DR3 to not significantly impact the two-body orbital solution. However, any unresolved companion with an intermediate orbital period ($\sim 4-75$y, \cite{gaia_collaboration_gaia_2022-1, Kervella2022}) can also be probed by the difference in proper motion it will have caused between Gaia and HIPPARCOS \citep{HIPPARCOS1997} single source proper motions. The proper motion anomaly between these catalogues for HIP 66074 is only $\chi^2 = 0.072$ \citep{Brandt2021}, where, for reference, a three sigma difference corresponds to $\chi^2 = 11.8$, leaving little room to hide for a third component. Even so, to test the scenario we extended our simulation of RV and astrometric measurements of binary systems, described in section \ref{sec:differentiating}, to encompass a third component. We also allowed the color, rotation speed and flux ratio of the components to vary significantly. Lastly, we took into account the slightly redder band pass of Gaia compared to HIRES. Depending on the configuration of stars, the band pass differences can amplify the relative brightness and thus the astrometric signal of two red binary stars relative to a hotter and bluer third component as much as a factor $\sim$\,1.9. The effects on RVs by the inclusion of a third component in the spectrum depends not only on the flux ratios of the components but also the widths of the absorption lines. A faster (slower) rotation of the third star compared to the two close components diminishes (increases) the impact of the third star on the combined spectrum, and thereby on the extracted RVs from that spectrum. The exact influence is further modified by the ratio of the rotational to orbital velocities at the epoch of observations and the RV extraction method, e.g.\ template matching, CCF, or the iodine method. Our simulations showed RV signals diminished up to $\sim$5-6 times (1.9 from the band pass and $\sim$2.5-3 from differences in flux and rotation) relative to the astrometric one but none of the configurations we tested were able to reproduce all the observables to a satisfying degree. %There is a limit to how much brighter the third, further-out star can be, because the combination of light from all the sources is determined by Gaia to be from a star with an effective temperature of $4161$K. %Secondly, a configuration where the light from e.g.\ two faint late-M dwarfs is mixed with that of a K-dwarf mutes not only the RV but also the astrometric signal of the M-dwarf orbit. 

Additionally, our simulations showed that the RVs were muted in an increasingly non-linear fashion as the influence of the third component increased. The resulting deformed shape of the RV curve meant worse fitting RVs at best and an unrecognizable RV curve at worst. We note that the HIRES observations employed the iodine method \citep[e.g.][]{Butler1996} and we tested here the effect of the third star on the CCFs and its RVs as measured by a Gaussian. We do not expect that the effect of the third star on the HIRES RVs to be exactly the same. 

%It was also difficult to find a configuration where this muting effect was many times stronger than the muting of the astrometric signal that would also occur (by  as shown in ).%, when the light from a static source is mixed in.

%\textit{Exotic object:}
%The above hypotheses all have significant issues, so further observations of this system will be of great interest. Based on the current data, the most likely explanation of the observation truly seem to be that some 
%very bright $0.445M_{\rm Jup}$ object is orbiting HIP 66074. This object could be some exotic or uncommon phenomenon, such as a planet with a highly anisotropic radiation pattern, such as a jet, or a very low-mass White Dwarf, or even a very low-mass black hole.

\paragraph{Nearly face-on orbit }
~An "easy" way to get a large difference in semi-amplitude between the astrometric and RV orbits is if the orbit of HIP 66074 is seen nearly face-on rather than edge-on. The orbital inclination of $i = 90.3 \pm 4.67 ^\circ$, as reported in Gaia DR3, does make this explanation seem unlikely. However, the one-dimensionality of the Gaia along-scan abscissa measurements for targets dimmer than G = 13 can give rise to degenerate solutions of the two-body models that are not necessarily close in parameter-space \citep{Sozzetti2022}. It may also be worth noting that the majority of exoplanet candidates lie above this exact magnitude threshold. The issue will be more prominent if the number of data points is not much larger than the number of free parameters of the model. As mentioned in section \ref{sec:intro}, the Gaia \texttt{nss\_two\_body\_orbit} solutions have 12 free parameters. In the case of HIP 66074, 38 measurements spread out over 28 distinct visibility periods\footnote{A visibility period is defined as "a group of observations separated from other groups by a gap of at least 4 days" as paraphrased from the Gaia online documentation, https://gea.esac.esa.int/archive/documentation/GDR3/ .} were used to calculate the astrometric solution. 28 visibility periods is in fact higher than the exoplanet candidate average, giving no obvious reason to distrust the quoted orbital inclination. Without access to the epoch data, a more in-depth analysis is not currently possible. \\

\subsection{Gaia DR3 1916454200349735680}

\begin{figure*}[h]
    \centering
    \includegraphics[width = \textwidth, trim= {3cm, 0cm 1cm 1cm}]{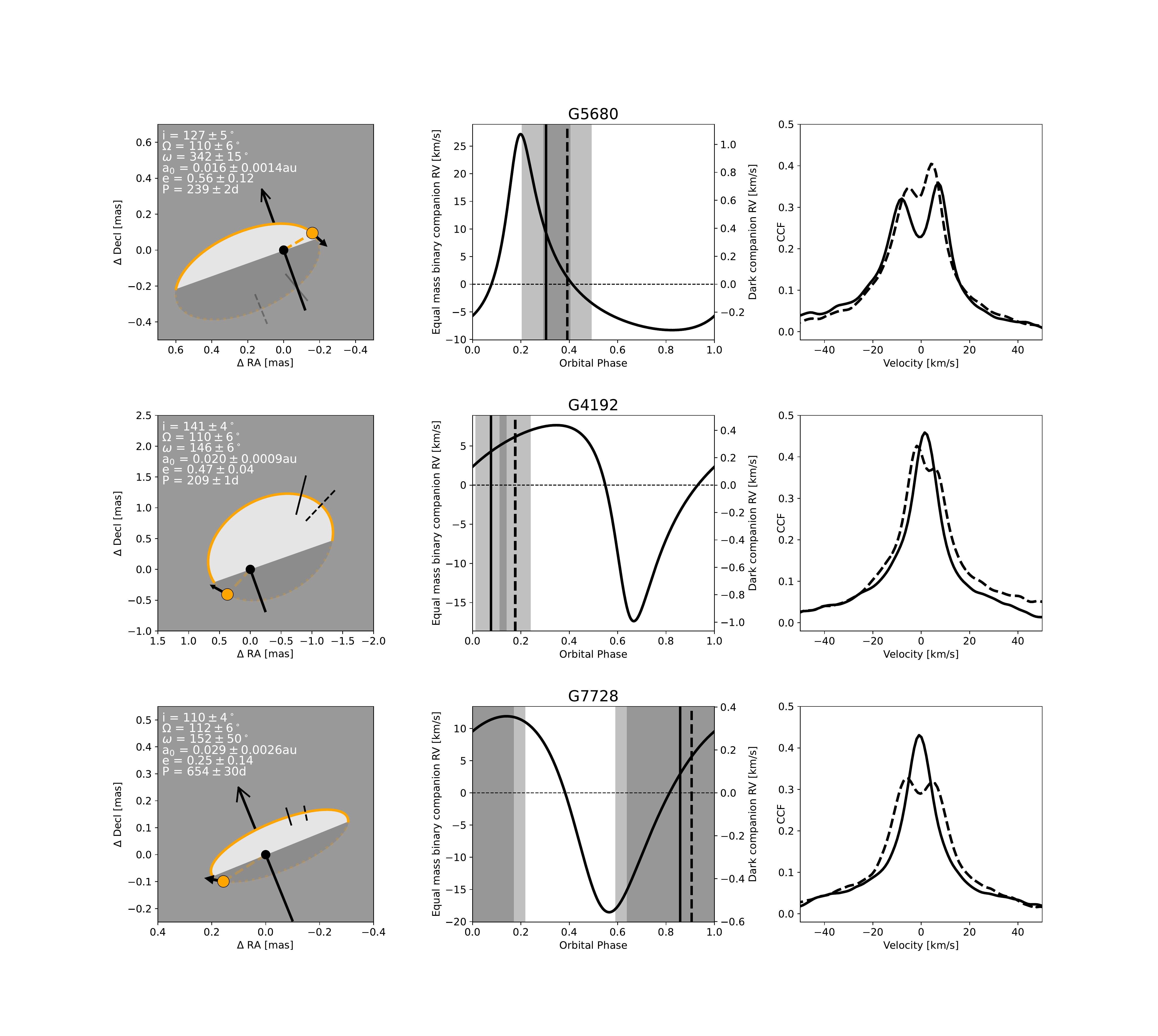}
    \caption{\textbf{NOT targets.} Each row corresponds to one of the three NOT fast track targets. These systems are, from top to bottom: Gaia DR3 052469973468984192, Gaia DR3 916454200349735680, and Gaia DR3 5122670101678217728. \textit{Left column:} The most likely orbits from the Gaia astrometric two-body solution are shown alongside the orbital parameters and their uncertainties. \textit{Middle column:} RV curve of the primary component as inferred from the Gaia orbital solution. The left y-axis scaling assumes the companion to have a mass equal to the primary. The right y-axis scaling assumes the companion to be dark. The two NOT observations are shown as solid and dotted lines. The gray shaded areas indicate the 1$\sigma$ uncertainties in orbital phase of these observations. The rectangles are darker where the uncertainties overlap. The large uncertainties in orbital phase timing are caused by the long baseline between the Gaia and NOT observations. \textit{Right column:} CCFs obtained from the FIES spectra, using line styles corresponding to those from the middle column.}
    \label{fig:Gaia5680}
\end{figure*}

Gaia DR3 1916454200349735680 (G5680) is the first of three systems we observed with the FIES spectrograph at the NOT. We obtained two spectra of the system three weeks apart. G5680 is characterized in Gaia as a K-dwarf and has a G magnitude of $10.9$. Its location in the HR-diagram is visibly elevated, see Fig.~\ref{fig:HR_candidates}. This elevation was the reason for including it in our small sample as it serves as a test of whether the location in an HR diagram can be used to gauge the likelihood of a binary false positive. The suspicion that this system was a binary was confirmed right away by inspecting the CCFs (Fig.~\ref{fig:Gaia5680}, top panel). The CCF of both spectra display a clear double peak structure indicating it is an SB2 binary. The similar heights and widths of the CCFs show that the two stars are very similar and that this system is a prototypical twin binary impostor. The difference in the separation between the CCF peaks implies a decrease in the radial component of the primary component's orbital velocity from $\sim 7.7$~km\,s$^{-1}$ to $\sim 5.8$~km\,s$^{-1}$ in the 21 days between our observations of the system. Such a decrease in RV fits with the first observation being at an orbital phase of around 0.92 and the second at 1, corresponding to 18.3 days. Here we assumed equal mass of the components equal to the $M_1$ \texttt{binary\_masses} value of $0.63 M_{\sun}$ and using Gaia's orbital inclination of $127^\circ$. The period and inclination of the \texttt{nss\_two\_body\_orbit} orbital solution are thus consistent with our observations.

\subsection{Gaia DR3 2052469973468984192}

Gaia DR3 2052469973468984192 (G4192), our second FIES target, is classified as an M-dwarf with a G magnitude of $10.7$. It is located right on the single star Main-Sequence in the HR-diagram (Fig.~\ref{fig:HR_candidates}), so the expectation would be that the small astrometric motion of the photocenter, $a_0 = 0.02$~au ($1.14$~mas) would originate from an orbiting Brown Dwarf with a mass of $22.9 \pm 1.6 M_{\rm Jup}$, as per equation \ref{eq:a0}. However, despite the first observation showing only a single CCF peak, this system is also a twin binary system, evident by the two-peaked CCF of the second observation obtained a few weeks later, see Fig.~\ref{fig:Gaia5680} middle panel. The separation between the CCF peaks increases from 0~km\,s$^{-1}$ to 4.6~km\,s$^{-1}$ between the two observations. Using the quoted inclination of $141^\circ$ and $M_1 = M_2 = 0.47M_{\sun}$, where the mass once again taken to be the \texttt{binary\_masses}$M_1$ mass, this RV separation would be expected to take $33.2$ days. The actual gap between the observations was 21 days. While not wildly different, we do not attempt to investigate whether this discrepancy is caused by an underestimate of $M_1$ or an overestimate of either the inclination or the orbital period. We do note however that $1/\sin141^\circ \approx 33.2 / 21$, or in other words, if the configuration of the orbit is edge-on it would exactly account for the discrepancy.
%Here we list the planet solution as well as a binary star solution. We note that the binary star solution is somewhat in tension with the location of the system on the single star main sequence. The near equal mass solution would shift the location up by XX in figure~\ref{fig:HR_candidates}. Indeed the first two observations are single peaked as expected for a single planet hosting star. Also the obtained RVs see table~\ref{tab:FIES_obs} are consistent with the planetary interpretation of the astrometric signal. \textbf{[I guess there is no point of any type of fit with two RVs?]} While the RVs cover only a small part of the orbit. We therefore give this system currently the tentative label of a planet hosting system.  

\subsection{DR3 5122670101678217728}

The third and last of our FIES Fast Track targets is the system Gaia DR3 5122670101678217728 (G7728). It is classified as a G-dwarf by Gaia and has a G magnitude of $8.9$. The observed photocenter semi-major axis is $a_0 = 0.029$~au ($0.28 \pm 0.03$ mas), which amounts to a companion mass of $15.6 \pm 1.7 M_{\rm Jup}$, assuming the companion is dark and using the mass estimate of the primary provided by Gaia. However, just like the two other FIES systems, the binary nature of the system is revealed by the double peak feature of the CCF obtained at a larger RV separation, see the bottom panel of Fig.~\ref{fig:Gaia5680}. The two peaks are extremely similar in height and width indicating that this system hosts two twin stars and not a star and a planet.
The separation between the CCF peaks corresponds to an RV increase of the primary of 6.4~km\,s$^{-1}$ over the 31 day period between our observations of this target. Using the same equal-mass assumption as the other systems and using $M_1 = 1.04 M_{\sun}$ and $i = 110^\circ$ corresponds to a $32.7$ day gap starting from an orbital phase of 0.4. Our observations are thus consistent with the orbital solution of this system.

%\begin{figure*}
%    \centering
%    \includegraphics[width = \textwidth, trim = {3cm, 1.2cm 3cm 3.8cm}]{figures/G7728.pdf}
%    \label{fig:Gaia7728}
%    \caption{\textbf{Gaia DR3 5122670101678217728.} }
%\end{figure*}

\section{Conclusion} \label{sec:conclusion}

With 9 of the 72 astrometric exoplanet candidates being previously confirmed exoplanets, Gaia's ability to detect photocenter movements small enough to find exoplanets is validated by DR3 and shows the current and future impact of the Gaia spacecraft on exoplanetary science. In this work we investigated the nature of six of these candidate systems using information contained in archival and newly obtained high resolution spectra. We found four systems, HD 68638, Gaia DR3 1916454200349735680, Gaia DR3 2052469973468984192, and Gaia DR3 5122670101678217728 to be false positives. %Their spectra consist each of two sets of spectral lines, shifting and merging in velocity space consistent with the Gaia DR3 orbital solution. 
The double-peaked CCFs of these systems show that they all consist of two nearly equal flux binaries. In the remaining two systems, HD 40503 and HIP 66074, the candidates are possibly exoplanets, although inconsistencies between the RV and astrometric solutions exist. The astrometric solutions imply in both cases an RV signal significantly larger than observed. Yet, all other orbital parameters are fully consistent between RV and astrometry. It is not clear what causes this, but a more face-on geometry of the orbits than implied by the astrometric solution could solve the disagreements.  %An overview of the current status of the Gaia astrometric candidate systems with ground-based data is shown in table \ref{tab:results}.

%This leaves the state of the released Gaia exoplanet candidates as follows:
%Of the 72 exoplanet candidates, ten are previously confirmed exoplanets: nine systems listed in the released candidate list\footnote{https://www.cosmos.esa.int/web/gaia/exoplanets} plus the exoplanet 2MASS J02192210-3925225b \citep{Artigau2015}. Another five are binary systems: BD+24 4592 (classified as a spectroscopic binary by \cite{Pourbaix2004}), HD 3321 (classified as a twin binary by \cite{Bonavita2022}) and our three NOT targets Gaia DR3 1916454200349735680, Gaia DR3 2052469973468984192 and Gaia DR3 5122670101678217728. All in all this leaves 57 new potential companions with masses $\leq 20M_{\rm Jup}$. 

We found 30 systems in the Gaia two-body orbital solutions with photocenter semi-major axes small enough to indicate a star-planet system that were not among the released exoplanet candidates. These systems are likely binary stars as per Gaia's flux ratio estimates, see Fig.~\ref{fig:HR_candidates}. These 30 systems along with the four false-positive exoplanets discussed in this paper highlight the difficulties of detecting exoplanets using astrometry. In this work we emphasize the need for spectra of high spectral resolution to supplement the astrometric data in order to reveal impostors by their double-lined nature or by the correlation of the FWHM with orbital phase. The high resolution is needed because the total wavelength shift of two sets of unresolved absorption lines from a binary system is dimmed in almost perfect proportion to the astrometric dimming, imitating an exoplanet signal in both data sets. The silver lining of the twin binaries issue is the fact that the similarity of the components makes it easier to identify the splitting in two of the absorption lines because the absorption lines have similar depths. %The six binary systems are our three NOT targets, Gaia DR3 1916454200349735680, Gaia DR3 2052469973468984192, Gaia DR3 5122670101678217728, HD 68638, plus the systems Gaia DR3 1878822452815621120, which was classified as a spectroscopic binary by \cite{Pourbaix2004}, and Gaia DR3 4901802507993393664, classified as a twin binary by \cite{Bonavita2022}. Another system, Gaia DR3 2047188847334279424 is listed as a visual binary with a separation of 34 mas, which is worth mentioning but likely not related to the 0.5 mas two-body solution by Gaia. We also note that two hosts of exoplanet candidates are appear to be white dwarfs: Gaia DR3 4698424845771339520 and Gaia DR3 6471102606408911360.

%\begin{longrotatetable}
\startlongtable
\begin{deluxetable*}{llcccccccccc}
\centerwidetable
\tablecaption{Apparent low-mass companions.} \label{tab:results}
\tablewidth{0pt}
\tablenum{1}
\tablehead{\colhead{Gaia DR3 id} &\colhead{Name} & \colhead{solution} & \colhead{Status} & \colhead{RV data} & \colhead{$M_1$}       & \colhead{$M_{2}$}     & \colhead{Flux ratio} & \colhead{G$_{Mag}$}& \colhead{Distance} & \colhead{References} \\
           \colhead{}      &\colhead{}             & \colhead{\_type}              & \colhead{}       & \colhead{}            &\colhead{($M_\sun$)}   & \colhead{($M_{J}$)} & \colhead{$F_2 / F_1$}           & \colhead{}         & \colhead{(pc)}     & \colhead{}} 
\startdata
 1035000055055287680 & HD 68638  & OTSV    & SB2       & Consistent   & $^{1}0.95^{+1.00}_{-0.82}$ & 31$^{+48}_{-14}$ & $\sim 1$     &  7.3 &  32.5 & 1,\,2,\,3            &  \\
 2884087104955208064 & HD 40503  & OTSV    & Unknown   & Inconsistent & $^{1}0.79^{+0.85}_{-0.57}$ & 5$^{+8}_{-3}$    & $[0 - 1]$    &  9.0 &  39.2 & 2,\,3                &  \\
 1712614124767394816 & HIP 66074 & OTSV    & Unknown   & Inconsistent & $^{1}0.71^{+0.76}_{-0.51}$ & 7$^{+11}_{-3}$   & $[0 - 1]$    &  9.7 &  35.4 & 2,\,3,\,4,\,5        &  \\
 1916454200349735680 &           & Orbital & SB2       & Consistent   & $^{1}0.63^{+0.68}_{-0.48}$ & 17$^{+26}_{-8}$  & $\sim 0.88$  & 10.9 &  37.1 & This work            &  \\
 2052469973468984192 &           & Orbital & SB2       & Consistent   & $^{1}0.47^{+0.52}_{-0.29}$ & 18$^{+27}_{-9}$  & $\sim 0.85$  & 10.7 &  17.4 & This work            &  \\
 5122670101678217728 & HD 12357  & Orbital & SB2       & Consistent   & $^{1}1.04^{+1.10}_{-0.97}$ & 21$^{+33}_{-10}$ & $\sim 1$     &  8.9 & 102.8 & This work            &  \\
 4062446910648807168 & HR 164604 & OTSV    & Exoplanet & No           & $^{2}0.75^{+0.79}_{-0.71}$ & 14$^{+24}_{-5}$  & $\sim 0$     &  9.3 &  40.0 & 12                   &  \\
 1594127865540229888 & HD 132406 & OTSV    & Exoplanet & Inconsistent & $^{2}0.97^{+1.01}_{-0.93}$ & 6$^{+11}_{-3}$   & $\sim 0$     &  8.3 &  70.5 & 4,\,8                &  \\
 2367734656180397952 & BD-17 63  & OTSV    & Exoplanet & Consistent   & $^{1}0.74^{+0.79}_{-0.56}$ & 4$^{+7}_{-2}$    & $\sim 0$     &  9.2 &  34.5 & 4,\,6                &  \\
 4745373133284418816 & HR 810    & OTS     & Exoplanet & Inconsistent & $^{1}1.09^{+1.15}_{-0.93}$ & 6$^{+10}_{-3}$   & $\sim 0$     &  5.3 &  17.4 & 4,\,5,\,11,\,13,\,14 &  \\
  637329067477530368 & HD 81040  & OTSV    & Exoplanet & Consistent   & $^{1}0.98^{+1.03}_{-0.68}$ & 8$^{+12}_{-4}$   & $\sim 0$     &  7.6 &  34.4 & 4,\,7                &  \\
 4901802507993393664 & HD 3221   & OTS     & SB2       & No           & $^{3}1$                    & 14               & $\sim 1$     &  9.1 &  44.4 & 17                   &  \\
 5855730584310531200 & HD 111232 & OTSV    & Exoplanet & Inconsistent & $^{1}0.93^{+0.99}_{-0.68}$ & 8$^{+13}_{-4}$   & $\sim 0$     &  7.4 &  28.9 & 4,\,9,\,10,\,11      &  \\
 2603090003484152064 & GJ 876    & OTSV    & Exoplanet & No           & $^{1}0.38^{+0.43}_{-0.21}$ & 4$^{+5}_{-2}$    & $\sim 0$     &  8.9 &   4.7 & 16                   &  \\
 4976894960284258048 & HD 142    & OTS     & Exoplanet & No           & $^{1}1.19^{+1.25}_{-1.06}$ & 7$^{+11}_{-3}$   & $\sim 0$     &  5.6 &  26.2 & 15                   &  \\
 6421118739093252224 & HD 175167 & OTSV    & Exoplanet & Inconsistent & $^{2}1.09^{+1.13}_{-1.05}$ & 10$^{+16}_{-5}$  & $\sim 0$     &  7.8 &  71.2 & 4,\,12               &  \\
 4764340705296117120 &           & Orbital & Unknown   & No           & $^{1}0.34^{+0.39}_{-0.18}$ & 13$^{+19}_{-6}$  & $[0 - 1]$    & 14.8 &  61.8 &                      &  \\
 1984587671751337600 &           & Orbital & Unknown   & No           & $^{1}0.36^{+0.41}_{-0.24}$ & 19$^{+29}_{-9}$  & $[0.34 - 1]$ & 14.2 &  54.0 &                      &  \\
 4392798821280870912 &           & Orbital & Unknown   & No           & $^{1}0.42^{+0.47}_{-0.26}$ & 15$^{+23}_{-7}$  & $[0.05 - 1]$ & 13.4 &  42.6 &                      &  \\
 5220375041387610880 &           & Orbital & Unknown   & No           & $^{1}0.74^{+0.79}_{-0.53}$ & 20$^{+31}_{-9}$  & $[0 - 1]$    & 12.8 & 183.6 &                      &  \\
 6418925831870553472 &           & Orbital & Unknown   & No           & $^{1}0.37^{+0.42}_{-0.20}$ & 15$^{+23}_{-8}$  & $[0 - 1]$    & 15.0 &  84.0 &                      &  \\
 2571855077162098944 &           & Orbital & Unknown   & No           & $^{1}0.54^{+0.59}_{-0.37}$ & 15$^{+23}_{-7}$  & $[0 - 1]$    & 14.5 & 114.6 &                      &  \\
 5036787935627755520 &           & Orbital & Unknown   & No           & $^{1}0.22^{+0.26}_{-0.10}$ & 12$^{+17}_{-6}$  & $[0.21 - 1]$ & 14.9 &  33.5 &                      &  \\
  932447162423519232 &           & ASSB1   & Unknown   & No           & $^{1}0.63^{+0.68}_{-0.48}$ & 17$^{+27}_{-8}$  & $[0.22 - 1]$ & 11.6 &  52.5 &                      &  \\
 5654515588409756160 &           & Orbital & Unknown   & No           & $^{1}0.26^{+0.31}_{-0.12}$ & 23$^{+34}_{-11}$ & $[0 - 1]$    & 16.4 &  88.4 &                      &  \\
 5271515801094390912 &           & Orbital & Unknown   & No           & $^{1}0.56^{+0.61}_{-0.39}$ & 21$^{+32}_{-10}$ & $[0 - 1]$    & 12.2 &  55.6 &                      &  \\
 5085864568417061120 &           & Orbital & Unknown   & No           & $^{1}0.46^{+0.51}_{-0.28}$ & 16$^{+24}_{-8}$  & $[0 - 1]$    & 13.7 &  55.8 &                      &  \\
  423297927866697088 &           & Orbital & Unknown   & No           & $^{1}0.41^{+0.46}_{-0.24}$ & 21$^{+32}_{-10}$ & $[0 - 1]$    & 14.8 &  79.2 &                      &  \\
 2845310284780420864 &           & Orbital & Unknown   & No           & $^{1}0.37^{+0.42}_{-0.20}$ & 10$^{+15}_{-5}$  & $[0 - 1]$    & 13.8 &  41.5 &                      &  \\
 4188996885011268608 &           & Orbital & Unknown   & No           & $^{1}0.23^{+0.28}_{-0.09}$ & 8$^{+12}_{-4}$   & $[0 - 1]$    & 13.5 &  17.8 &                      &  \\
 4842246017566495232 &           & Orbital & Unknown   & No           & $^{1}0.36^{+0.41}_{-0.19}$ & 8$^{+12}_{-4}$   & $[0 - 1]$    & 13.0 &  31.9 &                      &  \\
 1336053176328998144 &           & Orbital & Unknown   & No           & $^{1}0.44^{+0.49}_{-0.33}$ & 19$^{+29}_{-9}$  & $[0.48 - 1]$ & 14.3 &  83.2 &                      &  \\
 6081071334868194176 &           & Orbital & Unknown   & No           & $^{1}0.61^{+0.66}_{-0.43}$ & 20$^{+31}_{-10}$ & $[0 - 1]$    & 12.9 &  99.0 &                      &  \\
  405316961377489792 &           & Orbital & Unknown   & No           & $^{1}0.27^{+0.32}_{-0.13}$ & 22$^{+33}_{-10}$ & $[0 - 1]$    & 15.7 &  77.4 &                      &  \\
 6671454584430500864 &           & Orbital & Unknown   & No           & $^{1}0.57^{+0.62}_{-0.43}$ & 14$^{+24}_{-6}$  & $[0.17 - 1]$ & 13.9 & 112.4 &                      &  \\
 3621891774065137408 &           & Orbital & Unknown   & No           & $^{1}0.36^{+0.41}_{-0.22}$ & 19$^{+29}_{-9}$  & $[0.19 - 1]$ & 15.5 &  92.3 &                      &  \\
 4159075462792179456 &           & Orbital & Unknown   & No           & $^{1}0.35^{+0.40}_{-0.20}$ & 8$^{+13}_{-4}$   & $[0.07 - 1]$ & 14.9 &  64.4 &                      &  \\
 3665298981300771200 &           & Orbital & Unknown   & No           & $^{1}0.69^{+0.74}_{-0.54}$ & 20$^{+31}_{-9}$  & $[0.29 - 1]$ & 10.7 &  48.7 &                      &  \\
 4810127839810808576 &           & Orbital & Unknown   & No           & $^{1}0.55^{+0.60}_{-0.38}$ & 16$^{+24}_{-7}$  & $[0.07 - 1]$ & 13.0 &  61.4 &                      &  \\
 2998643469106143104 &           & Orbital & Unknown   & No           & $^{1}0.37^{+0.41}_{-0.20}$ & 20$^{+30}_{-10}$ & $[0 - 1]$    & 14.0 &  48.3 &                      &  \\
  557717892980808960 &           & Orbital & Unknown   & No           & $^{1}0.55^{+0.60}_{-0.37}$ & 8$^{+13}_{-4}$   & $[0 - 1]$    & 12.0 &  40.0 &                      &  \\
 6381440834777420928 &           & Orbital & Unknown   & No           & $^{1}0.28^{+0.33}_{-0.13}$ & 21$^{+32}_{-10}$ & $[0 - 1]$    & 15.1 &  60.6 &                      &  \\
 4596564611107874944 &           & Orbital & Unknown   & No           & $^{1}0.32^{+0.37}_{-0.17}$ & 19$^{+29}_{-9}$  & $[0.10 - 1]$ & 14.8 &  54.9 &                      &  \\
 2271703211828512896 &           & Orbital & Unknown   & No           & $^{1}0.16^{+0.21}_{-0.07}$ & 13$^{+20}_{-6}$  & $[0.30 - 1]$ & 16.4 &  42.3 &                      &  \\
 5612039087715504640 &           & Orbital & Unknown   & No           & $^{1}0.28^{+0.33}_{-0.13}$ & 14$^{+21}_{-7}$  & $[0 - 1]$    & 13.9 &  32.4 &                      &  \\
  834357565445682944 &           & Orbital & Unknown   & No           & $^{1}0.43^{+0.48}_{-0.25}$ & 17$^{+25}_{-8}$  & $[0 - 1]$    & 13.7 &  53.6 &                      &  \\
 5446516751833167744 &           & Orbital & Unknown   & No           & $^{1}0.27^{+0.32}_{-0.12}$ & 17$^{+25}_{-8}$  & $[0.01 - 1]$ & 16.0 &  70.4 &                      &  \\
  846058258950880384 &           & Orbital & Unknown   & No           & $^{1}0.60^{+0.65}_{-0.49}$ & 15$^{+24}_{-7}$  & $[0.44 - 1]$ & 13.9 & 143.4 &                      &  \\
 2117364283603705984 &           & Orbital & Unknown   & No           & $^{1}0.41^{+0.46}_{-0.29}$ & 15$^{+23}_{-7}$  & $[0.37 - 1]$ & 13.5 &  48.9 &                      &  \\
 2074815898041643520 &           & Orbital & Unknown   & No           & $^{1}0.36^{+0.41}_{-0.20}$ & 21$^{+32}_{-10}$ & $[0 - 1]$    & 13.2 &  41.3 &                      &  \\
 2047188847334279424 &           & OTS     & Unknown   & No           & $^{1}0.96^{+1.01}_{-0.82}$ & 14$^{+22}_{-7}$  & $[0 - 1]$    &  7.3 &  32.6 &                      &  \\
 6685861691447769600 &           & Orbital & Unknown   & No           & $^{1}0.50^{+0.55}_{-0.32}$ & 20$^{+31}_{-10}$ & $[0 - 1]$    & 14.3 &  99.6 &                      &  \\
 6079316686107743488 &           & Orbital & Unknown   & No           & $^{1}0.38^{+0.43}_{-0.21}$ & 20$^{+31}_{-10}$ & $[0 - 1]$    & 14.7 &  66.4 &                      &  \\
  430892357759527424 &           & Orbital & Unknown   & No           & $^{1}0.19^{+0.24}_{-0.08}$ & 7$^{+11}_{-3}$   & $[0.15 - 1]$ & 15.6 &  38.1 &                      &  \\
 1298992006611690112 &           & Orbital & Unknown   & No           & $^{1}0.52^{+0.57}_{-0.36}$ & 14$^{+21}_{-7}$  & $[0.11 - 1]$ & 13.6 &  73.9 &                      &  \\
 1878822452815621120 &           & OTS     & Unknown   & No           & $^{1}0.82^{+0.88}_{-0.61}$ & 16$^{+25}_{-8}$  & $[0 - 1]$    &  9.8 &  62.3 &                      &  \\
 1879554280883275136 &           & Orbital & Unknown   & No           & $^{1}0.43^{+0.48}_{-0.26}$ & 18$^{+27}_{-9}$  & $[0 - 1]$    & 14.3 &  79.3 &                      &  \\
 6471102606408911360 &           & Orbital & Unknown   & No           & $^{1}0.65^{+0.81}_{-0.49}$ & 23$^{+36}_{-11}$ & $[0 - 1]$    & 14.3 &  62.1 &                      &  \\
 5671384265738137984 &           & Orbital & Unknown   & No           & $^{1}0.15^{+0.20}_{-0.07}$ & 16$^{+24}_{-8}$  & $[0.27 - 1]$ & 15.7 &  28.3 &                      &  \\
 2104920835634141696 &           & Orbital & Unknown   & No           & $^{1}0.21^{+0.26}_{-0.08}$ & 13$^{+19}_{-6}$  & $[0 - 1]$    & 15.4 &  46.2 &                      &  \\
  373892712892466048 &           & Orbital & Unknown   & No           & $^{1}0.51^{+0.56}_{-0.33}$ & 17$^{+25}_{-8}$  & $[0 - 1]$    & 13.9 &  75.3 &                      &  \\
 5399010462168339456 &           & Orbital & Unknown   & No           & $^{1}0.47^{+0.52}_{-0.39}$ & 14$^{+22}_{-6}$  & $[0.65 - 1]$ & 13.8 &  79.6 &                      &  \\
 5618776310850226432 &           & Orbital & Unknown   & No           & $^{1}0.47^{+0.52}_{-0.29}$ & 21$^{+32}_{-10}$ & $[0 - 1]$    & 14.7 &  94.1 &                      &  \\
 6781298098147816192 &           & Orbital & Unknown   & No           & $^{1}0.32^{+0.37}_{-0.16}$ & 6$^{+10}_{-3}$   & $[0 - 1]$    & 14.4 &  48.1 &                      &  \\
 2050702366781291776 &           & Orbital & Unknown   & No           & $^{1}0.26^{+0.31}_{-0.13}$ & 18$^{+27}_{-9}$  & $[0.22 - 1]$ & 16.4 &  85.2 &                      &  \\
 1052042828882790016 &           & Orbital & Unknown   & No           & $^{1}0.47^{+0.52}_{-0.30}$ & 21$^{+32}_{-10}$ & $[0 - 1]$    & 14.0 &  78.1 &                      &  \\
  726588585356221568 &           & Orbital & Unknown   & No           & $^{1}0.27^{+0.32}_{-0.15}$ & 20$^{+30}_{-9}$  & $[0.28 - 1]$ & 15.6 &  67.8 &                      &  \\
 5486916932205092352 &           & Orbital & Unknown   & No           & $^{1}0.33^{+0.38}_{-0.17}$ & 11$^{+16}_{-6}$  & $[0 - 1]$    & 12.2 &  17.1 &                      &  \\
 5052449001298518528 &           & Orbital & Unknown   & No           & $^{1}0.47^{+0.52}_{-0.29}$ & 20$^{+31}_{-10}$ & $[0 - 1]$    & 13.6 &  59.2 &                      &  \\
  246890014559489792 &           & Orbital & Unknown   & No           & $^{1}0.36^{+0.41}_{-0.19}$ & 6$^{+9}_{-3}$    & $[0 - 1]$    & 14.4 &  52.5 &                      &  \\
 5055723587443420928 &           & Orbital & Unknown   & No           & $^{1}0.45^{+0.50}_{-0.27}$ & 19$^{+30}_{-9}$  & $[0 - 1]$    & 14.6 & 137.7 &                      &  \\
 3676303512147120512 &           & Orbital & Unknown   & No           & $^{1}0.23^{+0.28}_{-0.10}$ & 18$^{+27}_{-9}$  & $[0 - 1]$    & 14.8 &  38.7 &                      &  \\
 1059462676944293376 &           & Orbital & Unknown   & No           & $^{1}0.48^{+0.53}_{-0.35}$ & 16$^{+24}_{-8}$  & $[0.33 - 1]$ & 14.7 & 111.6 &                      &  \\
 5490183684330661504 &           & Orbital & Unknown   & No           & $^{1}0.14^{+0.19}_{-0.05}$ & 23$^{+36}_{-11}$ & $[0 - 1]$    & 17.7 &  72.4 &                      &  \\
 2277249663873880576 &           & Orbital & Unknown   & No           & $^{1}0.50^{+0.55}_{-0.32}$ & 11$^{+17}_{-5}$  & $[0 - 1]$    & 12.9 &  51.1 &                      &  \\
 2259699048817216256 &           & Orbital & Unknown   & No           & $^{1}0.62^{+0.67}_{-0.49}$ & 16$^{+24}_{-7}$  & $[0.31 - 1]$ & 13.1 & 105.9 &                      &  \\
 2259968811419624448 &           & Orbital & Unknown   & No           & $^{1}0.40^{+0.45}_{-0.23}$ & 22$^{+33}_{-10}$ & $[0 - 1]$    & 13.9 &  54.9 &                      &  \\
 4812716639938468992 &           & Orbital & Unknown   & No           & $^{1}0.31^{+0.36}_{-0.15}$ & 22$^{+33}_{-11}$ & $[0 - 1]$    & 14.7 &  48.2 &                      &  \\
 6521749994635476992 &           & Orbital & Unknown   & No           & $^{1}0.48^{+0.53}_{-0.30}$ & 15$^{+22}_{-7}$  & $[0 - 1]$    & 13.6 &  62.4 &                      &  \\
 1457486023639239296 &           & Orbital & Unknown   & No           & $^{1}0.64^{+0.69}_{-0.45}$ & 13$^{+21}_{-6}$  & $[0 - 1]$    & 11.9 &  73.8 &                      &  \\
 1462767459023424512 &           & Orbital & Unknown   & No           & $^{1}0.37^{+0.41}_{-0.19}$ & 8$^{+12}_{-4}$   & $[0 - 1]$    & 15.1 &  72.0 &                      &  \\
   73648110622521600 &           & Orbital & Unknown   & No           & $^{1}0.25^{+0.30}_{-0.11}$ & 20$^{+30}_{-10}$ & $[0 - 1]$    & 15.5 &  58.6 &                      &  \\
 5796338299045711232 &           & Orbital & Unknown   & No           & $^{1}0.18^{+0.23}_{-0.08}$ & 11$^{+17}_{-5}$  & $[0.37 - 1]$ & 14.8 &  25.8 &                      &  \\
 3937630969071148032 &           & Orbital & Unknown   & No           & $^{1}0.43^{+0.48}_{-0.26}$ & 20$^{+30}_{-10}$ & $[0 - 1]$    & 14.7 &  84.7 &                      &  \\
 2446599193562312320 &           & Orbital & Unknown   & No           & $^{1}0.50^{+0.55}_{-0.33}$ & 24$^{+37}_{-11}$ & $[0 - 1]$    & 14.7 & 129.1 &                      &  \\
 1862136504889464192 &           & Orbital & Unknown   & No           & $^{1}0.37^{+0.42}_{-0.21}$ & 16$^{+23}_{-8}$  & $[0 - 1]$    & 15.2 &  83.1 &                      &  \\
 3925216795598987264 &           & Orbital & Unknown   & No           & $^{1}0.40^{+0.45}_{-0.25}$ & 19$^{+29}_{-9}$  & $[0.16 - 1]$ & 14.8 &  75.0 &                      &  \\
 2824801747222539648 &           & Orbital & Unknown   & No           & $^{1}0.44^{+0.49}_{-0.27}$ & 19$^{+29}_{-9}$  & $[0 - 1]$    & 14.0 &  70.4 &                      &  \\
 6694115931396057728 &           & Orbital & Unknown   & No           & $^{1}0.33^{+0.38}_{-0.17}$ & 9$^{+14}_{-5}$   & $[0 - 1]$    & 13.4 &  38.1 &                      &  \\
 4698424845771339520 &           & OTS     & Unknown   & No           & $^{1}0.65^{+0.81}_{-0.49}$ & 9$^{+14}_{-4}$   & $[0 - 1]$    & 13.7 &   9.7 &                      &  \\
 4702845638429469056 &           & Orbital & Unknown   & No           & $^{1}0.43^{+0.47}_{-0.26}$ & 7$^{+10}_{-3}$   & $[0 - 1]$    & 13.9 &  68.2 &                      &  \\
 6677563745912843776 &           & Orbital & Unknown   & No           & $^{1}0.39^{+0.44}_{-0.21}$ & 23$^{+35}_{-11}$ & $[0 - 1]$    & 15.5 &  94.7 &                      &  \\
 6354671987249126784 &           & ASSB1   & Unknown   & No           & $^{1}0.69^{+0.75}_{-0.64}$ & 13$^{+21}_{-6}$  & $[0.12 - 1]$ &  9.5 &  32.5 &                      &  \\
  522135261462534528 &           & OTS     & Unknown   & No           & $^{1}1.09^{+1.15}_{-0.79}$ & 7$^{+11}_{-3}$   & $[0 - 1]$    &  6.4 &  27.0 &                      &  \\
 5375875638010549376 &           & Orbital & Unknown   & No           & $^{1}0.40^{+0.45}_{-0.23}$ & 14$^{+21}_{-6}$  & $[0 - 1]$    & 14.4 &  67.1 &                      &  \\
 1610837178107032192 &           & OTS     & Unknown   & No           & $^{1}1.13^{+1.19}_{-0.97}$ & 13$^{+21}_{-6}$  & $[0 - 1]$    &  8.2 &  73.7 &                      &  \\
 4963614887043956096 &           & OTS     & Unknown   & No           & $^{3}1$                    & 46               & $[0 - 1]$    & 15.0 &  40.2 &                      &  \\
 3120450116011961984 &           & Orbital & Unknown   & No           & $^{1}0.64^{+0.69}_{-0.49}$ & 16$^{+25}_{-8}$  & $[0.21 - 1]$ & 12.8 &  94.7 &                      &  \\
 4983571882081864960 &           & Orbital & Unknown   & No           & $^{1}0.47^{+0.52}_{-0.29}$ & 12$^{+19}_{-6}$  & $[0 - 1]$    & 13.3 &  55.5 &                      &  \\
 5236626338671861760 &           & Orbital & Unknown   & No           & $^{3}1$                    & 21               & $[0 - 1]$    & 13.9 &  51.3 &                      &  \\
 5437488554482255872 &           & Orbital & Unknown   & No           & $^{1}0.50^{+0.55}_{-0.37}$ & 18$^{+27}_{-8}$  & $[0.29 - 1]$ & 13.4 &  67.4 &                      &  \\
  198464052134353536 &           & Orbital & Unknown   & No           & $^{1}0.20^{+0.25}_{-0.09}$ & 11$^{+17}_{-5}$  & $[0.22 - 1]$ & 14.4 &  24.1 &                      &  \\
  \enddata
\tablecomments{ The 102 listed systems are the union of the following subsets: A) the released Gaia exoplanet candidates and B) all systems whose Gaia \texttt{nss\_two\_body\_orbit} implies $M_2 < 20M_J$ assuming a dark companion, regardless of the \texttt{binary\_masses} flux ratio value. The same systems are shown in Fig.~\ref{fig:HR_candidates} and Fig.~\ref{fig:overview}. The source of information for the mass of the primary component $M_1$ and its uncertainty, was in order of decreasing priority: $M_1$ from \texttt{binary\_masses}, "mass\_flame" from \texttt{astrophysical\_parameters} and $M_1 = 1M_\sun$, denoted by $1, 2$ and $3$ respectively in the corresponding column of the table. The top six systems of the table are those analyzed in this paper. The system name is shown systems previously studied in the literature. Whether RV data from ground-based spectroscopy is available for a system, and if so, whether it is consistent with the \texttt{nss\_two\_body\_orbit} solution and companion mass or not is listed in the "RV data" column. We define the data sets to be consistent if all the orbital parameters as well as the RV semi-amplitudes are within $3\sigma$ of one another. However, for our three systems with only two spectra each, only the orbital period and RV semi-amplitudes are tested for consistency. The distance column is 1$/$parallax, where the parallax is from the \texttt{gaia\_source} table. The solution\_type column refers to the \texttt{nss\_solution\_type} of the \texttt{nss\_two\_body\_orbit} table. OTS, OTSV, and ASSB1 refer to orbitalTargetedSearch, OrbitalTargetedSearchValidated and AstroSpectroSB1 respectively. References: (1) \cite{Busa2007}; (2) \cite{holl_gaia_2022}; (3) \cite{gaia_collaboration_gaia_2022-1}; (4) \cite{winn_joint_2022}; (5) \cite{Butler2017}; (6) \cite{Moutou2009}; (7) \cite{Sozzetti2006}; (8) \cite{DaSilva2007}; (9) \cite{Minniti2009}; (10) \cite{Mayor2004}; (11) \cite{Trifonov2020}; (12) \cite{Arriagada2010}; (13) \cite{Kurster2000}; (14) \cite{Naef2001}; (15) \cite{Wittenmyer2012}; (16) \cite{Marci2001}; (17) \cite{Bonavita2022}.}
\end{deluxetable*}
%\end{longrotatetable}

\acknowledgments
This work has made use of data from the European Space Agency (ESA) mission
{\it Gaia} (\url{https://www.cosmos.esa.int/gaia}), processed by the {\it Gaia}
Data Processing and Analysis Consortium (DPAC,
\url{https://www.cosmos.esa.int/web/gaia/dpac/consortium}). Funding for the DPAC
has been provided by national institutions, in particular the institutions
participating in the {\it Gaia} Multilateral Agreement.

We thank Berry Holl for valuable clarification about the Gaia exoplanet pipeline and the exoplanet sample.

We thank Emil Knudstrup for insightful discussions concerning RV measurements and FIES spectra in particular. 

Funding for the Stellar Astrophysics Centre is provided by The Danish National Research Foundation (Grant agreement no.: DNRF106).

We acknowledge the support from the Danish Council for Independent Research through a grant, No.2032-00230B.


%\vspace{5mm}
%\facilities{HST(STIS), Swift(XRT and UVOT), AAVSO, CTIO:1.3m,CTIO:1.5m,CXO}


%\software{astropy \citep{2013A&A...558A..33A,2018AJ....156..123A},  
%          Cloudy \citep{2013RMxAA..49..137F}, 
%          Source Extractor \citep{1996A&AS..117..393B}
%          }

\bibliography{GAIA_planet_candidates_2}

\begin{thebibliography}{}
\expandafter\ifx\csname natexlab\endcsname\relax\def\natexlab#1{#1}\fi
\providecommand{\url}[1]{\href{#1}{#1}}
\providecommand{\dodoi}[1]{doi:~\href{http://doi.org/#1}{\nolinkurl{#1}}}
\providecommand{\doeprint}[1]{\href{http://ascl.net/#1}{\nolinkurl{http://ascl.net/#1}}}
\providecommand{\doarXiv}[1]{\href{https://arxiv.org/abs/#1}{\nolinkurl{https://arxiv.org/abs/#1}}}

\bibitem[{HIP(1997)}]{HIPPARCOS1997}
 1997, ESA Special Publication, Vol. 1200, {The HIPPARCOS and TYCHO catalogues.
  Astrometric and photometric star catalogues derived from the ESA HIPPARCOS
  Space Astrometry Mission}

\bibitem[{{Arriagada} {et~al.}(2010){Arriagada}, {Butler}, {Minniti},
  {L{\'o}pez-Morales}, {Shectman}, {Adams}, {Boss}, \&
  {Chambers}}]{Arriagada2010}
{Arriagada}, P., {Butler}, R.~P., {Minniti}, D., {et~al.} 2010,
  \href{http://dx.doi.org/10.1088/0004-637X/711/2/1229}{\color{magenta}\apj},
  \href{https://ui.adsabs.harvard.edu/abs/2010ApJ...711.1229A}{\color{cyan}711},
  1229, \dodoi{10.1088/0004-637X/711/2/1229}

\bibitem[{{Bonavita} {et~al.}(2022){Bonavita}, {Gratton}, {Desidera},
  {Squicciarini}, {D'Orazi}, {Zurlo}, {Biller}, {Chauvin}, {Fontanive},
  {Janson}, {Messina}, {Menard}, {Meyer}, {Vigan}, {Avenhaus}, {Asensio
  Torres}, {Beuzit}, {Boccaletti}, {Bonnefoy}, {Brandner}, {Cantalloube},
  {Cheetham}, {Cudel}, {Daemgen}, {Delorme}, {Desgrange}, {Dominik}, {Engler},
  {Feautrier}, {Feldt}, {Galicher}, {Garufi}, {Gasparri}, {Ginski}, {Girard},
  {Grandjean}, {Hagelberg}, {Henning}, {Hunziker}, {Kasper}, {Keppler},
  {Lagadec}, {Lagrange}, {Langlois}, {Lannier}, {Lazzoni}, {Le Coroller},
  {Ligi}, {Lombart}, {Maire}, {Mazevet}, {Mesa}, {Mouillet}, {Moutou},
  {M{\"u}ller}, {Peretti}, {Perrot}, {Petrus}, {Potier}, {Ramos}, {Rickman},
  {Rouan}, {Salter}, {Samland}, {Schmidt}, {Sissa}, {Stolker}, {Szul{\'a}gyi},
  {Turatto}, {Udry}, \& {Wildi}}]{Bonavita2022}
{Bonavita}, M., {Gratton}, R., {Desidera}, S., {et~al.} 2022,
  \href{http://dx.doi.org/10.1051/0004-6361/202140510}{\color{magenta}\aap},
  \href{https://ui.adsabs.harvard.edu/abs/2022A&A...663A.144B}{\color{cyan}663},
  A144, \dodoi{10.1051/0004-6361/202140510}

\bibitem[{{Brandt}(2021)}]{Brandt2021}
{Brandt}, T.~D. 2021,
  \href{http://dx.doi.org/10.3847/1538-4365/abf93c}{\color{magenta}\apjs},
  \href{https://ui.adsabs.harvard.edu/abs/2021ApJS..254...42B}{\color{cyan}254},
  42, \dodoi{10.3847/1538-4365/abf93c}

\bibitem[{{Bus{\`a}} {et~al.}(2007){Bus{\`a}}, {Aznar Cuadrado}, {Terranegra},
  {Andretta}, \& {Gomez}}]{Busa2007}
{Bus{\`a}}, I., {Aznar Cuadrado}, R., {Terranegra}, L., {Andretta}, V., \&
  {Gomez}, M.~T. 2007,
  \href{http://dx.doi.org/10.1051/0004-6361:20065588}{\color{magenta}\aap},
  \href{https://ui.adsabs.harvard.edu/abs/2007A&A...466.1089B}{\color{cyan}466},
  1089, \dodoi{10.1051/0004-6361:20065588}

\bibitem[{{Butler} {et~al.}(1996){Butler}, {Marcy}, {Williams}, {McCarthy},
  {Dosanjh}, \& {Vogt}}]{Butler1996}
{Butler}, R.~P., {Marcy}, G.~W., {Williams}, E., {et~al.} 1996,
  \href{http://dx.doi.org/10.1086/133755}{\color{magenta}\pasp},
  \href{https://ui.adsabs.harvard.edu/abs/1996PASP..108..500B}{\color{cyan}108},
  500, \dodoi{10.1086/133755}

\bibitem[{{Butler} {et~al.}(2017){Butler}, {Vogt}, {Laughlin}, {Burt},
  {Rivera}, {Tuomi}, {Teske}, {Arriagada}, {Diaz}, {Holden}, \&
  {Keiser}}]{Butler2017}
{Butler}, R.~P., {Vogt}, S.~S., {Laughlin}, G., {et~al.} 2017,
  \href{http://dx.doi.org/10.3847/1538-3881/aa66ca}{\color{magenta}\aj},
  \href{https://ui.adsabs.harvard.edu/abs/2017AJ....153..208B}{\color{cyan}153},
  208, \dodoi{10.3847/1538-3881/aa66ca}

\bibitem[{{Curiel} {et~al.}(2022){Curiel}, {Ortiz-Le{\'o}n}, {Mioduszewski}, \&
  {Sanchez-Bermudez}}]{Salvador2022}
{Curiel}, S., {Ortiz-Le{\'o}n}, G.~N., {Mioduszewski}, A.~J., \&
  {Sanchez-Bermudez}, J. 2022,
  \href{http://dx.doi.org/10.3847/1538-3881/ac7c66}{\color{magenta}\aj},
  \href{https://ui.adsabs.harvard.edu/abs/2022AJ....164...93C}{\color{cyan}164},
  93, \dodoi{10.3847/1538-3881/ac7c66}

\bibitem[{{da Silva} {et~al.}(2007){da Silva}, {Udry}, {Bouchy}, {Moutou},
  {Mayor}, {Beuzit}, {Bonfils}, {Delfosse}, {Desort}, {Forveille}, {Galland},
  {H{\'e}brard}, {Lagrange}, {Loeillet}, {Lovis}, {Pepe}, {Perrier}, {Pont},
  {Queloz}, {Santos}, {S{\'e}gransan}, {Sivan}, {Vidal-Madjar}, \&
  {Zucker}}]{DaSilva2007}
{da Silva}, R., {Udry}, S., {Bouchy}, F., {et~al.} 2007,
  \href{http://dx.doi.org/10.1051/0004-6361:20077314}{\color{magenta}\aap},
  \href{https://ui.adsabs.harvard.edu/abs/2007A&A...473..323D}{\color{cyan}473},
  323, \dodoi{10.1051/0004-6361:20077314}

\bibitem[{{Gaia Collaboration} {et~al.}(2016){Gaia Collaboration}, {Prusti},
  {de Bruijne}, {Brown}, {Vallenari}, {Babusiaux}, {Bailer-Jones}, {Bastian},
  {Biermann}, {Evans}, {Eyer}, {Jansen}, {Jordi}, {Klioner}, {Lammers},
  {Lindegren}, {Luri}, {Mignard}, {Milligan}, {Panem}, {Poinsignon},
  {Pourbaix}, {Randich}, {Sarri}, {Sartoretti}, {Siddiqui}, {Soubiran},
  {Valette}, {van Leeuwen}, {Walton}, {Aerts}, {Arenou}, {Cropper}, {Drimmel},
  {H{\o}g}, {Katz}, {Lattanzi}, {O'Mullane}, {Grebel}, {Holland}, {Huc},
  {Passot}, {Bramante}, {Cacciari}, {Casta{\~n}eda}, {Chaoul}, {Cheek}, {De
  Angeli}, {Fabricius}, {Guerra}, {Hern{\'a}ndez}, {Jean-Antoine-Piccolo},
  {Masana}, {Messineo}, {Mowlavi}, {Nienartowicz}, {Ord{\'o}{\~n}ez-Blanco},
  {Panuzzo}, {Portell}, {Richards}, {Riello}, {Seabroke}, {Tanga},
  {Th{\'e}venin}, {Torra}, {Els}, {Gracia-Abril}, {Comoretto},
  {Garcia-Reinaldos}, {Lock}, {Mercier}, {Altmann}, {Andrae}, {Astraatmadja},
  {Bellas-Velidis}, {Benson}, {Berthier}, {Blomme}, {Busso}, {Carry},
  {Cellino}, {Clementini}, {Cowell}, {Creevey}, {Cuypers}, {Davidson}, {De
  Ridder}, {de Torres}, {Delchambre}, {Dell'Oro}, {Ducourant}, {Fr{\'e}mat},
  {Garc{\'\i}a-Torres}, {Gosset}, {Halbwachs}, {Hambly}, {Harrison}, {Hauser},
  {Hestroffer}, {Hodgkin}, {Huckle}, {Hutton}, {Jasniewicz}, {Jordan},
  {Kontizas}, {Korn}, {Lanzafame}, {Manteiga}, {Moitinho}, {Muinonen},
  {Osinde}, {Pancino}, {Pauwels}, {Petit}, {Recio-Blanco}, {Robin}, {Sarro},
  {Siopis}, {Smith}, {Smith}, {Sozzetti}, {Thuillot}, {van Reeven}, {Viala},
  {Abbas}, {Abreu Aramburu}, {Accart}, {Aguado}, {Allan}, {Allasia},
  {Altavilla}, {{\'A}lvarez}, {Alves}, {Anderson}, {Andrei}, {Anglada Varela},
  {Antiche}, {Antoja}, {Ant{\'o}n}, {Arcay}, {Atzei}, {Ayache}, {Bach},
  {Baker}, {Balaguer-N{\'u}{\~n}ez}, {Barache}, {Barata}, {Barbier}, {Barblan},
  {Baroni}, {Barrado y Navascu{\'e}s}, {Barros}, {Barstow}, {Becciani},
  {Bellazzini}, {Bellei}, {Bello Garc{\'\i}a}, {Belokurov}, {Bendjoya},
  {Berihuete}, {Bianchi}, {Bienaym{\'e}}, {Billebaud}, {Blagorodnova},
  {Blanco-Cuaresma}, {Boch}, {Bombrun}, {Borrachero}, {Bouquillon}, {Bourda},
  {Bouy}, {Bragaglia}, {Breddels}, {Brouillet}, {Br{\"u}semeister},
  {Bucciarelli}, {Budnik}, {Burgess}, {Burgon}, {Burlacu}, {Busonero}, {Buzzi},
  {Caffau}, {Cambras}, {Campbell}, {Cancelliere}, {Cantat-Gaudin}, {Carlucci},
  {Carrasco}, {Castellani}, {Charlot}, {Charnas}, {Charvet}, {Chassat},
  {Chiavassa}, {Clotet}, {Cocozza}, {Collins}, {Collins}, {Costigan}, {Crifo},
  {Cross}, {Crosta}, {Crowley}, {Dafonte}, {Damerdji}, {Dapergolas}, {David},
  {David}, {De Cat}, {de Felice}, {de Laverny}, {De Luise}, {De March}, {de
  Martino}, {de Souza}, {Debosscher}, {del Pozo}, {Delbo}, {Delgado},
  {Delgado}, {di Marco}, {Di Matteo}, {Diakite}, {Distefano}, {Dolding}, {Dos
  Anjos}, {Drazinos}, {Dur{\'a}n}, {Dzigan}, {Ecale}, {Edvardsson}, {Enke},
  {Erdmann}, {Escolar}, {Espina}, {Evans}, {Eynard Bontemps}, {Fabre},
  {Fabrizio}, {Faigler}, {Falc{\~a}o}, {Farr{\`a}s Casas}, {Faye}, {Federici},
  {Fedorets}, {Fern{\'a}ndez-Hern{\'a}ndez}, {Fernique}, {Fienga}, {Figueras},
  {Filippi}, {Findeisen}, {Fonti}, {Fouesneau}, {Fraile}, {Fraser}, {Fuchs},
  {Furnell}, {Gai}, {Galleti}, {Galluccio}, {Garabato}, {Garc{\'\i}a-Sedano},
  {Gar{\'e}}, {Garofalo}, {Garralda}, {Gavras}, {Gerssen}, {Geyer}, {Gilmore},
  {Girona}, {Giuffrida}, {Gomes}, {Gonz{\'a}lez-Marcos},
  {Gonz{\'a}lez-N{\'u}{\~n}ez}, {Gonz{\'a}lez-Vidal}, {Granvik}, {Guerrier},
  {Guillout}, {Guiraud}, {G{\'u}rpide}, {Guti{\'e}rrez-S{\'a}nchez}, {Guy},
  {Haigron}, {Hatzidimitriou}, {Haywood}, {Heiter}, {Helmi}, {Hobbs},
  {Hofmann}, {Holl}, {Holland}, {Hunt}, {Hypki}, {Icardi}, {Irwin}, {Jevardat
  de Fombelle}, {Jofr{\'e}}, {Jonker}, {Jorissen}, {Julbe}, {Karampelas},
  {Kochoska}, {Kohley}, {Kolenberg}, {Kontizas}, {Koposov}, {Kordopatis},
  {Koubsky}, {Kowalczyk}, {Krone-Martins}, {Kudryashova}, {Kull}, {Bachchan},
  {Lacoste-Seris}, {Lanza}, {Lavigne}, {Le Poncin-Lafitte}, {Lebreton},
  {Lebzelter}, {Leccia}, {Leclerc}, {Lecoeur-Taibi}, {Lemaitre}, {Lenhardt},
  {Leroux}, {Liao}, {Licata}, {Lindstr{\o}m}, {Lister}, {Livanou}, {Lobel},
  {L{\"o}ffler}, {L{\'o}pez}, {Lopez-Lozano}, {Lorenz}, {Loureiro},
  {MacDonald}, {Magalh{\~a}es Fernandes}, {Managau}, {Mann}, {Mantelet},
  {Marchal}, {Marchant}, {Marconi}, {Marie}, {Marinoni}, {Marrese},
  {Marschalk{\'o}}, {Marshall}, {Mart{\'\i}n-Fleitas}, {Martino}, {Mary},
  {Matijevi{\v{c}}}, {Mazeh}, {McMillan}, {Messina}, {Mestre}, {Michalik},
  {Millar}, {Miranda}, {Molina}, {Molinaro}, {Molinaro}, {Moln{\'a}r},
  {Moniez}, {Montegriffo}, {Monteiro}, {Mor}, {Mora}, {Morbidelli}, {Morel},
  {Morgenthaler}, {Morley}, {Morris}, {Mulone}, {Muraveva}, {Musella},
  {Narbonne}, {Nelemans}, {Nicastro}, {Noval}, {Ord{\'e}novic},
  {Ordieres-Mer{\'e}}, {Osborne}, {Pagani}, {Pagano}, {Pailler}, {Palacin},
  {Palaversa}, {Parsons}, {Paulsen}, {Pecoraro}, {Pedrosa}, {Pentik{\"a}inen},
  {Pereira}, {Pichon}, {Piersimoni}, {Pineau}, {Plachy}, {Plum}, {Poujoulet},
  {Pr{\v{s}}a}, {Pulone}, {Ragaini}, {Rago}, {Rambaux}, {Ramos-Lerate},
  {Ranalli}, {Rauw}, {Read}, {Regibo}, {Renk}, {Reyl{\'e}}, {Ribeiro},
  {Rimoldini}, {Ripepi}, {Riva}, {Rixon}, {Roelens}, {Romero-G{\'o}mez},
  {Rowell}, {Royer}, {Rudolph}, {Ruiz-Dern}, {Sadowski}, {Sagrist{\`a}
  Sell{\'e}s}, {Sahlmann}, {Salgado}, {Salguero}, {Sarasso}, {Savietto},
  {Schnorhk}, {Schultheis}, {Sciacca}, {Segol}, {Segovia}, {Segransan},
  {Serpell}, {Shih}, {Smareglia}, {Smart}, {Smith}, {Solano}, {Solitro},
  {Sordo}, {Soria Nieto}, {Souchay}, {Spagna}, {Spoto}, {Stampa}, {Steele},
  {Steidelm{\"u}ller}, {Stephenson}, {Stoev}, {Suess}, {S{\"u}veges}, {Surdej},
  {Szabados}, {Szegedi-Elek}, {Tapiador}, {Taris}, {Tauran}, {Taylor},
  {Teixeira}, {Terrett}, {Tingley}, {Trager}, {Turon}, {Ulla}, {Utrilla},
  {Valentini}, {van Elteren}, {Van Hemelryck}, {van Leeuwen}, {Varadi},
  {Vecchiato}, {Veljanoski}, {Via}, {Vicente}, {Vogt}, {Voss}, {Votruba},
  {Voutsinas}, {Walmsley}, {Weiler}, {Weingrill}, {Werner}, {Wevers},
  {Whitehead}, {Wyrzykowski}, {Yoldas}, {{\v{Z}}erjal}, {Zucker}, {Zurbach},
  {Zwitter}, {Alecu}, {Allen}, {Allende Prieto}, {Amorim},
  {Anglada-Escud{\'e}}, {Arsenijevic}, {Azaz}, {Balm}, {Beck}, {Bernstein},
  {Bigot}, {Bijaoui}, {Blasco}, {Bonfigli}, {Bono}, {Boudreault}, {Bressan},
  {Brown}, {Brunet}, {Bunclark}, {Buonanno}, {Butkevich}, {Carret}, {Carrion},
  {Chemin}, {Ch{\'e}reau}, {Corcione}, {Darmigny}, {de Boer}, {de Teodoro}, {de
  Zeeuw}, {Delle Luche}, {Domingues}, {Dubath}, {Fodor}, {Fr{\'e}zouls},
  {Fries}, {Fustes}, {Fyfe}, {Gallardo}, {Gallegos}, {Gardiol}, {Gebran},
  {Gomboc}, {G{\'o}mez}, {Grux}, {Gueguen}, {Heyrovsky}, {Hoar}, {Iannicola},
  {Isasi Parache}, {Janotto}, {Joliet}, {Jonckheere}, {Keil}, {Kim},
  {Klagyivik}, {Klar}, {Knude}, {Kochukhov}, {Kolka}, {Kos}, {Kutka}, {Lainey},
  {LeBouquin}, {Liu}, {Loreggia}, {Makarov}, {Marseille}, {Martayan},
  {Martinez-Rubi}, {Massart}, {Meynadier}, {Mignot}, {Munari}, {Nguyen},
  {Nordlander}, {Ocvirk}, {O'Flaherty}, {Olias Sanz}, {Ortiz}, {Osorio},
  {Oszkiewicz}, {Ouzounis}, {Palmer}, {Park}, {Pasquato}, {Peltzer}, {Peralta},
  {P{\'e}turaud}, {Pieniluoma}, {Pigozzi}, {Poels}, {Prat}, {Prod'homme},
  {Raison}, {Rebordao}, {Risquez}, {Rocca-Volmerange}, {Rosen}, {Ruiz-Fuertes},
  {Russo}, {Sembay}, {Serraller Vizcaino}, {Short}, {Siebert}, {Silva},
  {Sinachopoulos}, {Slezak}, {Soffel}, {Sosnowska}, {Strai{\v{z}}ys}, {ter
  Linden}, {Terrell}, {Theil}, {Tiede}, {Troisi}, {Tsalmantza}, {Tur},
  {Vaccari}, {Vachier}, {Valles}, {Van Hamme}, {Veltz}, {Virtanen}, {Wallut},
  {Wichmann}, {Wilkinson}, {Ziaeepour}, \& {Zschocke}}]{2016GaiaMission}
{Gaia Collaboration}, {Prusti}, T., {de Bruijne}, J.~H.~J., {et~al.} 2016,
  \href{http://dx.doi.org/10.1051/0004-6361/201629272}{\color{magenta}\aap},
  \href{https://ui.adsabs.harvard.edu/abs/2016A&A...595A...1G}{\color{cyan}595},
  A1, \dodoi{10.1051/0004-6361/201629272}

\bibitem[{{Gaia Collaboration} {et~al.}(2022{\natexlab{a}}){Gaia
  Collaboration}, {Vallenari}, {Brown}, {Prusti}, {de Bruijne}, {Arenou},
  {Babusiaux}, {Biermann}, {Creevey}, {Ducourant}, {Evans}, {Eyer}, {Guerra},
  {Hutton}, {Jordi}, {Klioner}, {Lammers}, {Lindegren}, {Luri}, {Mignard},
  {Panem}, {Pourbaix}, {Randich}, {Sartoretti}, {Soubiran}, {Tanga}, {Walton},
  {Bailer-Jones}, {Bastian}, {Drimmel}, {Jansen}, {Katz}, {Lattanzi}, {van
  Leeuwen}, {Bakker}, {Cacciari}, {Casta{\~n}eda}, {De Angeli}, {Fabricius},
  {Fouesneau}, {Fr{\'e}mat}, {Galluccio}, {Guerrier}, {Heiter}, {Masana},
  {Messineo}, {Mowlavi}, {Nicolas}, {Nienartowicz}, {Pailler}, {Panuzzo},
  {Riclet}, {Roux}, {Seabroke}, {Sordo{\o}rcit}, {Th{\'e}venin},
  {Gracia-Abril}, {Portell}, {Teyssier}, {Altmann}, {Andrae}, {Audard},
  {Bellas-Velidis}, {Benson}, {Berthier}, {Blomme}, {Burgess}, {Busonero},
  {Busso}, {C{\'a}novas}, {Carry}, {Cellino}, {Cheek}, {Clementini},
  {Damerdji}, {Davidson}, {de Teodoro}, {Nu{\~n}ez Campos}, {Delchambre},
  {Dell'Oro}, {Esquej}, {Fern{\'a}ndez-Hern{\'a}ndez}, {Fraile}, {Garabato},
  {Garc{\'\i}a-Lario}, {Gosset}, {Haigron}, {Halbwachs}, {Hambly}, {Harrison},
  {Hern{\'a}ndez}, {Hestroffer}, {Hodgkin}, {Holl}, {Jan{\ss}en}, {Jevardat de
  Fombelle}, {Jordan}, {Krone-Martins}, {Lanzafame}, {L{\"o}ffler}, {Marchal},
  {Marrese}, {Moitinho}, {Muinonen}, {Osborne}, {Pancino}, {Pauwels},
  {Recio-Blanco}, {Reyl{\'e}}, {Riello}, {Rimoldini}, {Roegiers}, {Rybizki},
  {Sarro}, {Siopis}, {Smith}, {Sozzetti}, {Utrilla}, {van Leeuwen}, {Abbas},
  {{\'A}brah{\'a}m}, {Abreu Aramburu}, {Aerts}, {Aguado}, {Ajaj},
  {Aldea-Montero}, {Altavilla}, {{\'A}lvarez}, {Alves}, {Anders}, {Anderson},
  {Anglada Varela}, {Antoja}, {Baines}, {Baker}, {Balaguer-N{\'u}{\~n}ez},
  {Balbinot}, {Balog}, {Barache}, {Barbato}, {Barros}, {Barstow},
  {Bartolom{\'e}}, {Bassilana}, {Bauchet}, {Becciani}, {Bellazzini},
  {Berihuete}, {Bernet}, {Bertone}, {Bianchi}, {Binnenfeld}, {Blanco-Cuaresma},
  {Blazere}, {Boch}, {Bombrun}, {Bossini}, {Bouquillon}, {Bragaglia},
  {Bramante}, {Breedt}, {Bressan}, {Brouillet}, {Brugaletta}, {Bucciarelli},
  {Burlacu}, {Butkevich}, {Buzzi}, {Caffau}, {Cancelliere}, {Cantat-Gaudin},
  {Carballo}, {Carlucci}, {Carnerero}, {Carrasco}, {Casamiquela}, {Castellani},
  {Castro-Ginard}, {Chaoul}, {Charlot}, {Chemin}, {Chiaramida}, {Chiavassa},
  {Chornay}, {Comoretto}, {Contursi}, {Cooper}, {Cornez}, {Cowell}, {Crifo},
  {Cropper}, {Crosta}, {Crowley}, {Dafonte}, {Dapergolas}, {David}, {David},
  {de Laverny}, {De Luise}, {De March}, {De Ridder}, {de Souza}, {de Torres},
  {del Peloso}, {del Pozo}, {Delbo}, {Delgado}, {Delisle}, {Demouchy},
  {Dharmawardena}, {Di Matteo}, {Diakite}, {Diener}, {Distefano}, {Dolding},
  {Edvardsson}, {Enke}, {Fabre}, {Fabrizio}, {Faigler}, {Fedorets}, {Fernique},
  {Fienga}, {Figueras}, {Fournier}, {Fouron}, {Fragkoudi}, {Gai},
  {Garcia-Gutierrez}, {Garcia-Reinaldos}, {Garc{\'\i}a-Torres}, {Garofalo},
  {Gavel}, {Gavras}, {Gerlach}, {Geyer}, {Giacobbe}, {Gilmore}, {Girona},
  {Giuffrida}, {Gomel}, {Gomez}, {Gonz{\'a}lez-N{\'u}{\~n}ez},
  {Gonz{\'a}lez-Santamar{\'\i}a}, {Gonz{\'a}lez-Vidal}, {Granvik}, {Guillout},
  {Guiraud}, {Guti{\'e}rrez-S{\'a}nchez}, {Guy}, {Hatzidimitriou}, {Hauser},
  {Haywood}, {Helmer}, {Helmi}, {Sarmiento}, {Hidalgo}, {Hilger},
  {H{\l}adczuk}, {Hobbs}, {Holland}, {Huckle}, {Jardine}, {Jasniewicz},
  {Jean-Antoine Piccolo}, {Jim{\'e}nez-Arranz}, {Jorissen}, {Juaristi
  Campillo}, {Julbe}, {Karbevska}, {Kervella}, {Khanna}, {Kontizas},
  {Kordopatis}, {Korn}, {K{\'o}sp{\'a}l}, {Kostrzewa-Rutkowska},
  {Kruszy{\'n}ska}, {Kun}, {Laizeau}, {Lambert}, {Lanza}, {Lasne}, {Le
  Campion}, {Lebreton}, {Lebzelter}, {Leccia}, {Leclerc}, {Lecoeur-Taibi},
  {Liao}, {Licata}, {Lindstr{\o}m}, {Lister}, {Livanou}, {Lobel}, {Lorca},
  {Loup}, {Madrero Pardo}, {Magdaleno Romeo}, {Managau}, {Mann}, {Manteiga},
  {Marchant}, {Marconi}, {Marcos}, {Marcos Santos}, {Mar{\'\i}n Pina},
  {Marinoni}, {Marocco}, {Marshall}, {Polo}, {Mart{\'\i}n-Fleitas}, {Marton},
  {Mary}, {Masip}, {Massari}, {Mastrobuono-Battisti}, {Mazeh}, {McMillan},
  {Messina}, {Michalik}, {Millar}, {Mints}, {Molina}, {Molinaro}, {Moln{\'a}r},
  {Monari}, {Mongui{\'o}}, {Montegriffo}, {Montero}, {Mor}, {Mora},
  {Morbidelli}, {Morel}, {Morris}, {Muraveva}, {Murphy}, {Musella}, {Nagy},
  {Noval}, {Oca{\~n}a}, {Ogden}, {Ordenovic}, {Osinde}, {Pagani}, {Pagano},
  {Palaversa}, {Palicio}, {Pallas-Quintela}, {Panahi}, {Payne-Wardenaar},
  {Pe{\~n}alosa Esteller}, {Penttil{\"a}}, {Pichon}, {Piersimoni}, {Pineau},
  {Plachy}, {Plum}, {Poggio}, {Pr{\v{s}}a}, {Pulone}, {Racero}, {Ragaini},
  {Rainer}, {Raiteri}, {Rambaux}, {Ramos}, {Ramos-Lerate}, {Re Fiorentin},
  {Regibo}, {Richards}, {Rios Diaz}, {Ripepi}, {Riva}, {Rix}, {Rixon},
  {Robichon}, {Robin}, {Robin}, {Roelens}, {Rogues}, {Rohrbasser},
  {Romero-G{\'o}mez}, {Rowell}, {Royer}, {Ruz Mieres}, {Rybicki}, {Sadowski},
  {S{\'a}ez N{\'u}{\~n}ez}, {Sagrist{\`a} Sell{\'e}s}, {Sahlmann}, {Salguero},
  {Samaras}, {Sanchez Gimenez}, {Sanna}, {Santove{\~n}a}, {Sarasso},
  {Schultheis}, {Sciacca}, {Segol}, {Segovia}, {S{\'e}gransan}, {Semeux},
  {Shahaf}, {Siddiqui}, {Siebert}, {Siltala}, {Silvelo}, {Slezak}, {Slezak},
  {Smart}, {Snaith}, {Solano}, {Solitro}, {Souami}, {Souchay}, {Spagna},
  {Spina}, {Spoto}, {Steele}, {Steidelm{\"u}ller}, {Stephenson}, {S{\"u}veges},
  {Surdej}, {Szabados}, {Szegedi-Elek}, {Taris}, {Taylo}, {Teixeira},
  {Tolomei}, {Tonello}, {Torra}, {Torra}, {Torralba Elipe}, {Trabucchi},
  {Tsounis}, {Turon}, {Ulla}, {Unger}, {Vaillant}, {van Dillen}, {van Reeven},
  {Vanel}, {Vecchiato}, {Viala}, {Vicente}, {Voutsinas}, {Weiler}, {Wevers},
  {Wyrzykowski}, {Yoldas}, {Yvard}, {Zhao}, {Zorec}, {Zucker}, \&
  {Zwitter}}]{2022DR3}
{Gaia Collaboration}, {Vallenari}, A., {Brown}, A.~G.~A., {et~al.}
  2022{\natexlab{a}},
  \href{http://dx.doi.org/10.48550/arXiv.2208.00211}{\color{magenta}arXiv
  e-prints}, arXiv:2208.00211, \dodoi{10.48550/arXiv.2208.00211}

\bibitem[{{Gaia Collaboration} {et~al.}(2022{\natexlab{b}}){Gaia
  Collaboration}, {Arenou}, {Babusiaux}, {Barstow}, {Faigler}, {Jorissen},
  {Kervella}, {Mazeh}, {Mowlavi}, {Panuzzo}, {Sahlmann}, {Shahaf}, {Sozzetti},
  {Bauchet}, {Damerdji}, {Gavras}, {Giacobbe}, {Gosset}, {Halbwachs}, {Holl},
  {Lattanzi}, {Leclerc}, {Morel}, {Pourbaix}, {Re Fiorentin}, {Sadowski},
  {S{\'e}gransan}, {Siopis}, {Teyssier}, {Zwitter}, {Planquart}, {Brown},
  {Vallenari}, {Prusti}, {de Bruijne}, {Biermann}, {Creevey}, {Ducourant},
  {Evans}, {Eyer}, {Guerra}, {Hutton}, {Jordi}, {Klioner}, {Lammers},
  {Lindegren}, {Luri}, {Mignard}, {Panem}, {Randich}, {Sartoretti}, {Soubiran},
  {Tanga}, {Walton}, {Bailer-Jones}, {Bastian}, {Drimmel}, {Jansen}, {Katz},
  {van Leeuwen}, {Bakker}, {Cacciari}, {Casta{\~n}eda}, {De Angeli},
  {Fabricius}, {Fouesneau}, {Fr{\'e}mat}, {Galluccio}, {Guerrier}, {Heiter},
  {Masana}, {Messineo}, {Nicolas}, {Nienartowicz}, {Pailler}, {Riclet}, {Roux},
  {Seabroke}, {Sordo}, {Th{\'e}venin}, {Gracia-Abril}, {Portell}, {Altmann},
  {Andrae}, {Audard}, {Bellas-Velidis}, {Benson}, {Berthier}, {Blomme},
  {Burgess}, {Busonero}, {Busso}, {C{\'a}novas}, {Carry}, {Cellino}, {Cheek},
  {Clementini}, {Davidson}, {de Teodoro}, {Nu{\~n}ez Campos}, {Delchambre},
  {Dell'Oro}, {Esquej}, {Fern{\'a}ndez-Hern{\'a}ndez}, {Fraile}, {Garabato},
  {Garc{\'\i}a-Lario}, {Haigron}, {Hambly}, {Harrison}, {Hern{\'a}ndez},
  {Hestroffer}, {Hodgkin}, {Jan{\ss}en}, {Jevardat de Fombelle}, {Jordan},
  {Krone-Martins}, {Lanzafame}, {L{\"o}ffler}, {Marchal}, {Marrese},
  {Moitinho}, {Muinonen}, {Osborne}, {Pancino}, {Pauwels}, {Recio-Blanco},
  {Reyl{\'e}}, {Riello}, {Rimoldini}, {Roegiers}, {Rybizki}, {Sarro}, {Smith},
  {Utrilla}, {van Leeuwen}, {Abbas}, {{\'A}brah{\'a}m}, {Abreu Aramburu},
  {Aerts}, {Aguado}, {Ajaj}, {Aldea-Montero}, {Altavilla}, {{\'A}lvarez},
  {Alves}, {Anders}, {Anderson}, {Anglada Varela}, {Antoja}, {Baines}, {Baker},
  {Balaguer-N{\'u}{\~n}ez}, {Balbinot}, {Balog}, {Barache}, {Barbato},
  {Barros}, {Bartolom{\'e}}, {Bassilana}, {Becciani}, {Bellazzini},
  {Berihuete}, {Bernet}, {Bertone}, {Bianchi}, {Binnenfeld}, {Blanco-Cuaresma},
  {Blazere}, {Boch}, {Bombrun}, {Bossini}, {Bouquillon}, {Bragaglia},
  {Bramante}, {Breedt}, {Bressan}, {Brouillet}, {Brugaletta}, {Bucciarelli},
  {Burlacu}, {Butkevich}, {Buzzi}, {Caffau}, {Cancelliere}, {Cantat-Gaudin},
  {Carballo}, {Carlucci}, {Carnerero}, {Carrasco}, {Casamiquela}, {Castellani},
  {Castro-Ginard}, {Chaoul}, {Charlot}, {Chemin}, {Chiaramida}, {Chiavassa},
  {Chornay}, {Comoretto}, {Contursi}, {Cooper}, {Cornez}, {Cowell}, {Crifo},
  {Cropper}, {Crosta}, {Crowley}, {Dafonte}, {Dapergolas}, {David}, {de
  Laverny}, {De Luise}, {De March}, {De Ridder}, {de Souza}, {de Torres}, {del
  Peloso}, {del Pozo}, {Delbo}, {Delgado}, {Delisle}, {Demouchy},
  {Dharmawardena}, {Diakite}, {Diener}, {Distefano}, {Dolding}, {Enke},
  {Fabre}, {Fabrizio}, {Fedorets}, {Fernique}, {Figueras}, {Fournier},
  {Fouron}, {Fragkoudi}, {Gai}, {Garcia-Gutierrez}, {Garcia-Reinaldos},
  {Garc{\'\i}a-Torres}, {Garofalo}, {Gavel}, {Gerlach}, {Geyer}, {Gilmore},
  {Girona}, {Giuffrida}, {Gomel}, {Gomez}, {Gonz{\'a}lez-N{\'u}{\~n}ez},
  {Gonz{\'a}lez-Santamar{\'\i}a}, {Gonz{\'a}lez-Vidal}, {Granvik}, {Guillout},
  {Guiraud}, {Guti{\'e}rrez-S{\'a}nchez}, {Guy}, {Hatzidimitriou}, {Hauser},
  {Haywood}, {Helmer}, {Helmi}, {Sarmiento}, {Hidalgo}, {H{\l}adczuk}, {Hobbs},
  {Holland}, {Huckle}, {Jardine}, {Jasniewicz}, {Jean-Antoine Piccolo},
  {Jim{\'e}nez-Arranz}, {Juaristi Campillo}, {Julbe}, {Karbevska}, {Khanna},
  {Kordopatis}, {Korn}, {K{\'o}sp{\'a}l}, {Kostrzewa-Rutkowska},
  {Kruszy{\'n}ska}, {Kun}, {Laizeau}, {Lambert}, {Lanza}, {Lasne}, {Le
  Campion}, {Lebreton}, {Lebzelter}, {Leccia}, {Lecoeur-Taibi}, {Liao},
  {Licata}, {Lindstr{\o}m}, {Lister}, {Livanou}, {Lobel}, {Lorca}, {Loup},
  {Madrero Pardo}, {Magdaleno Romeo}, {Managau}, {Mann}, {Manteiga},
  {Marchant}, {Marconi}, {Marcos}, {Marcos Santos}, {Mar{\'\i}n Pina},
  {Marinoni}, {Marocco}, {Marshall}, {Polo}, {Mart{\'\i}n-Fleitas}, {Marton},
  {Mary}, {Masip}, {Massari}, {Mastrobuono-Battisti}, {McMillan}, {Messina},
  {Michalik}, {Millar}, {Mints}, {Molina}, {Molinaro}, {Moln{\'a}r}, {Monari},
  {Mongui{\'o}}, {Montegriffo}, {Montero}, {Mor}, {Mora}, {Morbidelli},
  {Morris}, {Muraveva}, {Murphy}, {Musella}, {Nagy}, {Noval}, {Oca{\~n}a},
  {Ogden}, {Ordenovic}, {Osinde}, {Pagani}, {Pagano}, {Palaversa}, {Palicio},
  {Pallas-Quintela}, {Panahi}, {Payne-Wardenaar}, {Pe{\~n}alosa Esteller},
  {Penttil{\"a}}, {Pichon}, {Piersimoni}, {Pineau}, {Plachy}, {Plum}, {Poggio},
  {Pr{\v{s}}a}, {Pulone}, {Racero}, {Ragaini}, {Rainer}, {Raiteri}, {Ramos},
  {Ramos-Lerate}, {Regibo}, {Richards}, {Rios Diaz}, {Ripepi}, {Riva}, {Rix},
  {Rixon}, {Robichon}, {Robin}, {Robin}, {Roelens}, {Rogues}, {Rohrbasser},
  {Romero-G{\'o}mez}, {Rowell}, {Royer}, {Ruz Mieres}, {Rybicki}, {S{\'a}ez
  N{\'u}{\~n}ez}, {Sagrist{\`a} Sell{\'e}s}, {Salguero}, {Samaras}, {Sanchez
  Gimenez}, {Sanna}, {Santove{\~n}a}, {Sarasso}, {Schultheis}, {Sciacca},
  {Segol}, {Segovia}, {Semeux}, {Siddiqui}, {Siebert}, {Siltala}, {Silvelo},
  {Slezak}, {Slezak}, {Smart}, {Snaith}, {Solano}, {Solitro}, {Souami},
  {Souchay}, {Spagna}, {Spina}, {Spoto}, {Steele}, {Steidelm{\"u}ller},
  {Stephenson}, {S{\"u}veges}, {Surdej}, {Szabados}, {Szegedi-Elek}, {Taris},
  {Taylor}, {Teixeira}, {Tolomei}, {Tonello}, {Torra}, {Torra}, {Torralba
  Elipe}, {Trabucchi}, {Tsounis}, {Turon}, {Ulla}, {Unger}, {Vaillant}, {van
  Dillen}, {van Reeven}, {Vanel}, {Vecchiato}, {Viala}, {Vicente}, {Voutsinas},
  {Weiler}, {Wevers}, {Wyrzykowski}, {Yoldas}, {Yvard}, {Zhao}, {Zorec}, \&
  {Zucker}}]{gaia_collaboration_gaia_2022-1}
{Gaia Collaboration}, {Arenou}, F., {Babusiaux}, C., {et~al.}
  2022{\natexlab{b}},
  \href{http://dx.doi.org/10.48550/arXiv.2206.05595}{\color{magenta}arXiv
  e-prints}, arXiv:2206.05595, \dodoi{10.48550/arXiv.2206.05595}

\bibitem[{Halbwachs(2009)}]{halbwachs_local_2009}
Halbwachs, J.~L. 2009,
  \href{http://dx.doi.org/10.1111/j.1365-2966.2009.14406.x}{\color{magenta}Monthly
  Notices of the Royal Astronomical Society}, 394, 1075,
  \dodoi{10.1111/j.1365-2966.2009.14406.x}

\bibitem[{{Halbwachs} {et~al.}(2022){Halbwachs}, {Pourbaix}, {Arenou},
  {Galluccio}, {Guillout}, {Bauchet}, {Marchal}, {Sadowski}, \&
  {Teyssier}}]{halbwachs_gaia_2022}
{Halbwachs}, J.-L., {Pourbaix}, D., {Arenou}, F., {et~al.} 2022,
  \href{http://dx.doi.org/10.48550/arXiv.2206.05726}{\color{magenta}arXiv
  e-prints}, arXiv:2206.05726, \dodoi{10.48550/arXiv.2206.05726}

\bibitem[{{Holl} {et~al.}(2022{\natexlab{a}}){Holl}, {Fabricius}, {Portell},
  {Lindegren}, {Panuzzo}, {Bernet}, {Casta{\~n}eda}, {Jevardat de Fombelle},
  {Audard}, {Ducourant}, {Harrison}, {Evans}, {Busso}, {Sozzetti}, {Gosset},
  {Arenou}, {De Angeli}, {Riello}, {Eyer}, {Rimoldini}, {Gavras}, {Mowlavi},
  {Nienartowicz}, {Lecoeur-Ta{\"\i}bi}, {Garc{\'\i}a-Lario}, \&
  {Pourbaix}}]{holl_gaia_2022}
{Holl}, B., {Fabricius}, C., {Portell}, J., {et~al.} 2022{\natexlab{a}},
  \href{http://dx.doi.org/10.48550/arXiv.2212.11971}{\color{magenta}arXiv
  e-prints}, arXiv:2212.11971, \dodoi{10.48550/arXiv.2212.11971}

\bibitem[{{Holl} {et~al.}(2022{\natexlab{b}}){Holl}, {Fabricius}, {Portell},
  {Lindegren}, {Panuzzo}, {Bernet}, {Casta{\~n}eda}, {Jevardat de Fombelle},
  {Audard}, {Ducourant}, {Harrison}, {Evans}, {Busso}, {Sozzetti}, {Gosset},
  {Arenou}, {De Angeli}, {Riello}, {Eyer}, {Rimoldini}, {Gavras}, {Mowlavi},
  {Nienartowicz}, {Lecoeur-Ta{\"\i}bi}, {Garc{\'\i}a-Lario}, \&
  {Pourbaix}}]{Sozzetti2022}
---. 2022{\natexlab{b}},
  \href{http://dx.doi.org/10.48550/arXiv.2212.11971}{\color{magenta}arXiv
  e-prints}, arXiv:2212.11971, \dodoi{10.48550/arXiv.2212.11971}

\bibitem[{{Kervella} {et~al.}(2022){Kervella}, {Arenou}, \&
  {Th{\'e}venin}}]{Kervella2022}
{Kervella}, P., {Arenou}, F., \& {Th{\'e}venin}, F. 2022,
  \href{http://dx.doi.org/10.1051/0004-6361/202142146}{\color{magenta}\aap},
  \href{https://ui.adsabs.harvard.edu/abs/2022A&A...657A...7K}{\color{cyan}657},
  A7, \dodoi{10.1051/0004-6361/202142146}

\bibitem[{{K{\"u}rster} {et~al.}(2000){K{\"u}rster}, {Endl}, {Els}, {Hatzes},
  {Cochran}, {D{\"o}bereiner}, \& {Dennerl}}]{Kurster2000}
{K{\"u}rster}, M., {Endl}, M., {Els}, S., {et~al.} 2000, \aap,
  \href{https://ui.adsabs.harvard.edu/abs/2000A&A...353L..33K}{\color{cyan}353},
  L33

\bibitem[{{Marcy} {et~al.}(2001){Marcy}, {Butler}, {Fischer}, {Vogt},
  {Lissauer}, \& {Rivera}}]{Marci2001}
{Marcy}, G.~W., {Butler}, R.~P., {Fischer}, D., {et~al.} 2001,
  \href{http://dx.doi.org/10.1086/321552}{\color{magenta}\apj},
  \href{https://ui.adsabs.harvard.edu/abs/2001ApJ...556..296M}{\color{cyan}556},
  296, \dodoi{10.1086/321552}

\bibitem[{{Mayor} {et~al.}(2004){Mayor}, {Udry}, {Naef}, {Pepe}, {Queloz},
  {Santos}, \& {Burnet}}]{Mayor2004}
{Mayor}, M., {Udry}, S., {Naef}, D., {et~al.} 2004,
  \href{http://dx.doi.org/10.1051/0004-6361:20034250}{\color{magenta}\aap},
  \href{https://ui.adsabs.harvard.edu/abs/2004A&A...415..391M}{\color{cyan}415},
  391, \dodoi{10.1051/0004-6361:20034250}

\bibitem[{{Minniti} {et~al.}(2009){Minniti}, {Butler}, {L{\'o}pez-Morales},
  {Shectman}, {Adams}, {Arriagada}, {Boss}, \& {Chambers}}]{Minniti2009}
{Minniti}, D., {Butler}, R.~P., {L{\'o}pez-Morales}, M., {et~al.} 2009,
  \href{http://dx.doi.org/10.1088/0004-637X/693/2/1424}{\color{magenta}\apj},
  \href{https://ui.adsabs.harvard.edu/abs/2009ApJ...693.1424M}{\color{cyan}693},
  1424, \dodoi{10.1088/0004-637X/693/2/1424}

\bibitem[{{Moutou} {et~al.}(2009){Moutou}, {Mayor}, {Lo Curto}, {Udry},
  {Bouchy}, {Benz}, {Lovis}, {Naef}, {Pepe}, {Queloz}, \&
  {Santos}}]{Moutou2009}
{Moutou}, C., {Mayor}, M., {Lo Curto}, G., {et~al.} 2009,
  \href{http://dx.doi.org/10.1051/0004-6361:200810941}{\color{magenta}\aap},
  \href{https://ui.adsabs.harvard.edu/abs/2009A&A...496..513M}{\color{cyan}496},
  513, \dodoi{10.1051/0004-6361:200810941}

\bibitem[{{Naef} {et~al.}(2001){Naef}, {Mayor}, {Pepe}, {Queloz}, {Santos},
  {Udry}, \& {Burnet}}]{Naef2001}
{Naef}, D., {Mayor}, M., {Pepe}, F., {et~al.} 2001,
  \href{http://dx.doi.org/10.1051/0004-6361:20010841}{\color{magenta}\aap},
  \href{https://ui.adsabs.harvard.edu/abs/2001A&A...375..205N}{\color{cyan}375},
  205, \dodoi{10.1051/0004-6361:20010841}

\bibitem[{Perryman {et~al.}(2014)Perryman, Hartman, Bakos, \&
  Lindegren}]{perryman_astrometric_2014}
Perryman, M., Hartman, J., Bakos, G.~A., \& Lindegren, L. 2014,
  \href{http://dx.doi.org/10.1088/0004-637X/797/1/14}{\color{magenta}The
  Astrophysical Journal}, 797, 14, \dodoi{10.1088/0004-637X/797/1/14}

\bibitem[{{Sozzetti} {et~al.}(2006){Sozzetti}, {Udry}, {Zucker}, {Torres},
  {Beuzit}, {Latham}, {Mayor}, {Mazeh}, {Naef}, {Perrier}, {Queloz}, \&
  {Sivan}}]{Sozzetti2006}
{Sozzetti}, A., {Udry}, S., {Zucker}, S., {et~al.} 2006,
  \href{http://dx.doi.org/10.1051/0004-6361:20054303}{\color{magenta}\aap},
  \href{https://ui.adsabs.harvard.edu/abs/2006A&A...449..417S}{\color{cyan}449},
  417, \dodoi{10.1051/0004-6361:20054303}

\bibitem[{{Trifonov} {et~al.}(2020){Trifonov}, {Tal-Or}, {Zechmeister},
  {Kaminski}, {Zucker}, \& {Mazeh}}]{Trifonov2020}
{Trifonov}, T., {Tal-Or}, L., {Zechmeister}, M., {et~al.} 2020,
  \href{http://dx.doi.org/10.1051/0004-6361/201936686}{\color{magenta}\aap},
  \href{https://ui.adsabs.harvard.edu/abs/2020A&A...636A..74T}{\color{cyan}636},
  A74, \dodoi{10.1051/0004-6361/201936686}

\bibitem[{{Winn}(2022)}]{winn_joint_2022}
{Winn}, J.~N. 2022,
  \href{http://dx.doi.org/10.3847/1538-3881/ac9126}{\color{magenta}\aj},
  \href{https://ui.adsabs.harvard.edu/abs/2022AJ....164..196W}{\color{cyan}164},
  196, \dodoi{10.3847/1538-3881/ac9126}

\bibitem[{{Wittenmyer} {et~al.}(2012){Wittenmyer}, {Horner}, {Tuomi}, {Salter},
  {Tinney}, {Butler}, {Jones}, {O'Toole}, {Bailey}, {Carter}, {Jenkins},
  {Zhang}, {Vogt}, \& {Rivera}}]{Wittenmyer2012}
{Wittenmyer}, R.~A., {Horner}, J., {Tuomi}, M., {et~al.} 2012,
  \href{http://dx.doi.org/10.1088/0004-637X/753/2/169}{\color{magenta}\apj},
  \href{https://ui.adsabs.harvard.edu/abs/2012ApJ...753..169W}{\color{cyan}753},
  169, \dodoi{10.1088/0004-637X/753/2/169}

\end{thebibliography}
\bibliographystyle{aasjournal-hyperref}

\end{document}